\documentclass[journal,submit,pdftex,moreauthors]{Definitions/mdpi} 

\usepackage{comment}

\preto{\abstractkeywords}{\nolinenumbers}

\firstpage{1} 
\pubvolume{1}
\issuenum{1}
\articlenumber{0}
\pubyear{2024}
\copyrightyear{2024}
\datereceived{ } 
\daterevised{ } 
\dateaccepted{ } 
\datepublished{ } 
\hreflink{https://doi.org/} 

\Title{Spectropolarimetry for Discerning Geometry and Structure in Circumstellar Media of Hot Massive Stars}

\TitleCitation{Spectropolarimetry for Discerning Geometry and Structure in the Atmospheres and Circumstellar Media of Hot Massive Stars}

\Author{Richard Ignace $^{1,\dagger,\ddagger}$\orcidA{}, Kenneth G.\ Gayley $^{2}$, Roberto Casini$^{3}$, Paul Scowen$^{4}$, Christiana Erba$^{5}$, Jeremy Drake$^{6}$}

\AuthorNames{Richard Ignace, Kenneth Gayley, Roberto Cassini, Paul Scowen, Christiana Erba, and Jeremy Drake}

\AuthorCitation{Ignace, R.; Gayley, K.; Casini, R., Scowen, P., Erba, C., Drake, J.}

\address{%
$^{1}$ \quad Physics \& Astronomy, East Tennessee State University, Johnson City, TN, USA; ignace@etsu.edu\\
$^{2}$ \quad Physics \& Astronomy, University of Iowa, Iowa City, IA 52242, USA; kenneth-gayley@uiowa.edu\\
$^{3}$ \quad NSF NCAR High Altitude Observatory, P.O. Box 3000, Boulder, CO 80307-3000, USA \\
$^{4}$ \quad NASA Goddard Space Flight Center, Exoplanets and Stellar Astrophysics Lab, Greenbelt, MD 20771, USA\\
$^{5}$ \quad Space Telescope Science Institute, 3700 San Martin Drive, Baltimore, MD 21218, USA\\
$^{6}$ \quad Lockheed Martin, 3251 Hanover St, Palo Alto, CA, 94304, USA \\
}

\corres{Correspondence: ignace@etsu.edu}

\secondnote{All authors contributed to this work.}

\abstract{
Spectropolarimetric techniques are a mainstay of astrophysical
inquiry, ranging from Solar System objects to the Cosmic Background
Radiation.  This review highlights applications of stellar polarimetry
for massive hot stars, particularly in the context of ultraviolet
(UV) spaceborne missions.  The prevalence of binarity in the massive
star population and uncertainties regarding the degree of rotational
criticality among hot stars raises important questions about stellar
interactions, interior structure, and even the lifetimes of
evolutionary phases.  These uncertainties have consequences for
stellar population synthesis calculations.  Spectropolarimetry is
a key tool for extracting information about stellar and binary
geometries. We review methodologies involving electron scattering
in circumstellar envelopes; gravity darkening from rapid rotation;
spectral line effects including the (a) ``line effect'', (b) \"{O}hman
effect, and (c) Hanle effect; and the imprint of interstellar
polarization on measurements.  Finally, we describe the {\em Polstar}
UV spectropolarimetric SMEX mission concept as one means for employing
these diagnostics to clarify the state of high rotation and its
impacts for massive stars.
}


\begin{document}

\section{Introduction}

Massive stars are hot, luminous, UV-bright stars
with short lifetimes that explosively terminate to leave behind remnant compact objects \citep{2012ARA&A..50..107L}.
They are integral to understanding the history and futures of galaxy evolution, as massive stars are principal sites for nucleosynthesis and the distribution of metals throughout galaxies, which is important for planet formation and life as we know it \citep{2013ARA&A..51..457N}.  Their stories are relevant to gravitational wave
detections as progenitors of stellar mass black holes
\citep[e.g.,][]{2010ApJ...714.1217B}.

However, the specifics of massive star evolution are muddled by uncertainties surrounding binary interactions and the frequency and consequences of fast rotation
\citep[e.g.,][]{2022ARA&A..60..203V}.  It is now clear that massive
stars are routinely born into binaries, triples, or even higher multiples
\citep{2012Sci...337..444S, 2023A&A...678A..60K}.  Moreover, as a fraction of the rotational breakup speed,  
nearly all main sequence massive stars are fast rotating (above $\sim$30\%), with some
at near-critical rates (above 90\%).  Those that are slowly
rotating are likely magnetic \citep{2009MNRAS.392.1022U}. Binarity
combined with the incidence of runaway stars \citep{2019A&A...624A..66R}
and high rotation rates \citep{1996ApJ...463..737P, 2010ApJ...722..605H}
suggests that massive stars undergo considerable interactions,
such as mass transfer, common envelope evolution, and even mergers
\citep[e.g.,][]{2024arXiv240711680N}.  Such events involve the
exchange of angular momentum, including orbital evolution and stellar spin up \citep{1992ApJ...391..246P, 2020A&A...640A..16T}.  It remains unclear how much spin-up occurs, how angular momentum is transported throughout stellar interiors, and how much is conserved or lost to the system over the course of mass exchange.

\begin{figure}[t]
\begin{adjustwidth}{-\extralength}{-3cm}
\centering
\includegraphics[width=12 cm]{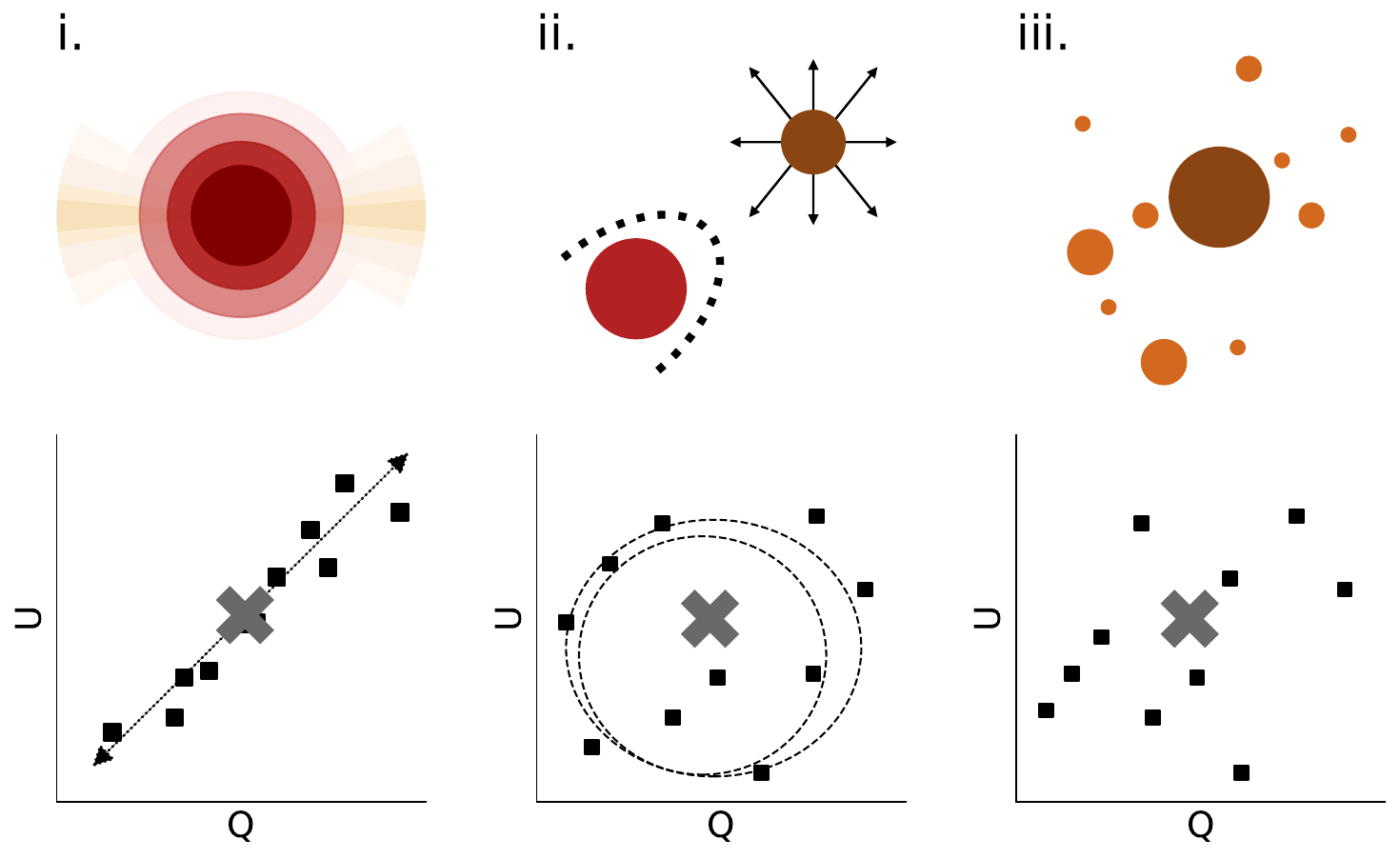}
\end{adjustwidth}
\caption{Categories i, ii, and iii of variable linear polarization from stellar sources as traced by Q-U diagrams.  Cat.~i is for a fixed PA on the sky where variability in the geometry or optical depth leads to linear variations between Q and U.  Cat.~ii shows loop behavior for cyclic variations, here using a binary star as example.  Finally, Cat.~iii displays a scatter diagram in Q-U for stochastic variability, shown here as an example of wind clumps.  In each case, the effect of ISP is signified by a vector translation of the respective temporal ensembles indicated by the ``X''.
\label{fig1}}
\end{figure}   
\unskip

 Progressing toward the resolution of these uncertainties will require techniques that can characterize the inherently non-spherical geometries involved. Spectropolarimetry is ideal for discerning geometry of spatially unresolved sources as characterized through the Stokes I, Q, U, and V parameters.  Circular polarization with Stokes V has proven successful in measuring the surface magnetism of massive stars \citep{2016MNRAS.456....2W, 2017A&A...599A..66S}; while linear polarimetry from Q and U of the scattered light probes geometry, making use of wavelength-dependent effects (e.g., spectral features) and time variability (e.g.,  perspective changes via phase monitoring).

An overview of relevant spectropolarimetric diagnostics is given in Section~\ref{sec2}, including continuum polarimetry and spectral line effects.  A case study involving an application of the Hanle effect to magnetospheric wind channeling is presented in Section~\ref{sec3}.
How interstellar polarization (ISP) affects the ability to extract science results for spectrpolarimetric data is described in Section~\ref{sec4}.  The {\em Polstar} mission concept for a new space telescope with a UV spectropolarimeter is detailed in Section~\ref{sec5}.

\section{Spectropolarimetric Techniques} \label{sec2}

Polarization is measured using the Stokes~I, Q, U, and V parameters \citep[e.g.,][]{1960ratr.book.....C}.
Stokes~I refers to the total measure of light.  Stokes~V refers to circular polarization.  Finally, Stokes~Q and U characterize the linear polarization.  With linear polarimetry the focus of this contribution, we ignore V hereafter.  

With detector measurements expressed as fluxes, 
$F_I$, $F_Q$, and $F_U$, the relative polarizations are given by
$q=F_Q/F_I$, and $u=F_U/F_I$. Then, the degree of net polarization
$p$ and polarization position angle $\psi_{\rm p}$ (or PA) become

\begin{eqnarray}
p & = & \sqrt{q^2+u^2},\\
\tan 2\psi_{\rm p} & = & u/q,
\end{eqnarray}

\noindent where wavelength dependence is implied.

Figure~\ref{fig1} illustrates scenarios involving variable polarization with time (note that similar trends can arise as chromatic effects, where points are for wavelengths instead of time samples).  We introduce 3 categories i, ii, and iii from left to right.  The upper panels are for physical scenarios; the lower ones illustrate temporal behavior in Q-U diagrams.  In each case the large ``X'' represents the interstellar polarization (ISP).  The categories are respectively (i) for stellar variability that maintains a fixed PA on the sky, in this case a disk that changes opening angle or opacity; (ii) for evolving orientation, in this case for a binary leading to cyclical loops; and (iii) for random variations, in this case represented by a stochastically clumpy wind.  As demonstrated by Figure~\ref{fig1}, the Q-U diagram is a valuable tool, whether for time or wavelength varying effects, to probe geometry in spatially unresolved stars.  

\begin{figure}[t]
\begin{adjustwidth}{-\extralength}{-3cm}
\centering
\includegraphics[width=11cm]{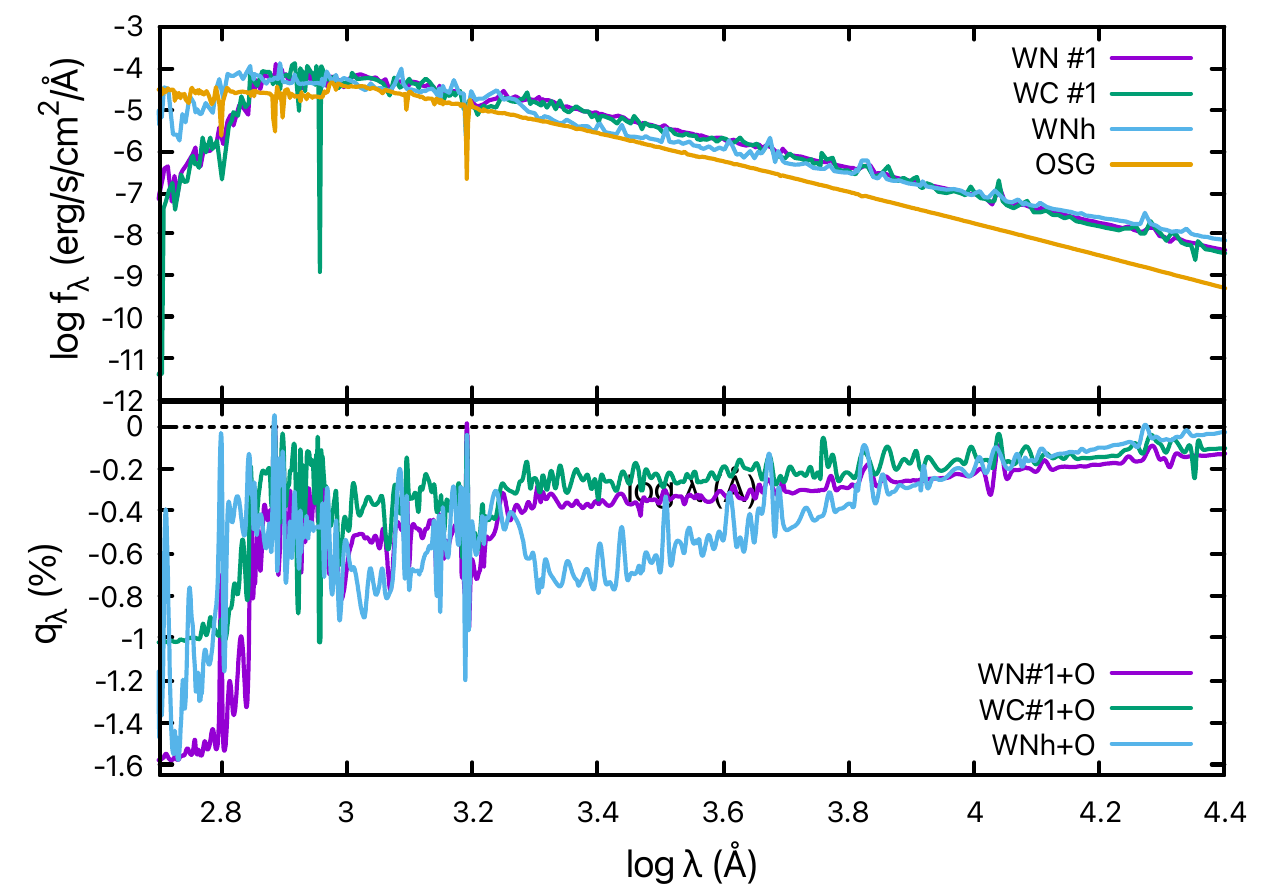}
\end{adjustwidth}
\caption{The polarization from electron scattering for WR+O colliding wind models.  Three cases are shown, all using an O supergiant (OSG) combined with:  a WN star, a WC star, and a WNh star, as described in the text.  The upper panel shows the combined flux assuming stars are at a distance of 10~pc. The lower panel is $q_\lambda$ in percent, with negative values signifying that the net polarization is perpendicular to the line of centers joining the two stars.  The polarization is not gray because the net $q_\lambda$ consists of a flux-weighted sum.  The wavelengths span from about 500~\AA\ (EUV) to 2.5~$\mu$m (NIR).
\label{fig2a}}
\end{figure}   
\unskip

\begin{figure}[t]
\begin{adjustwidth}{-\extralength}{-3cm}
\centering
\includegraphics[width=11 cm]{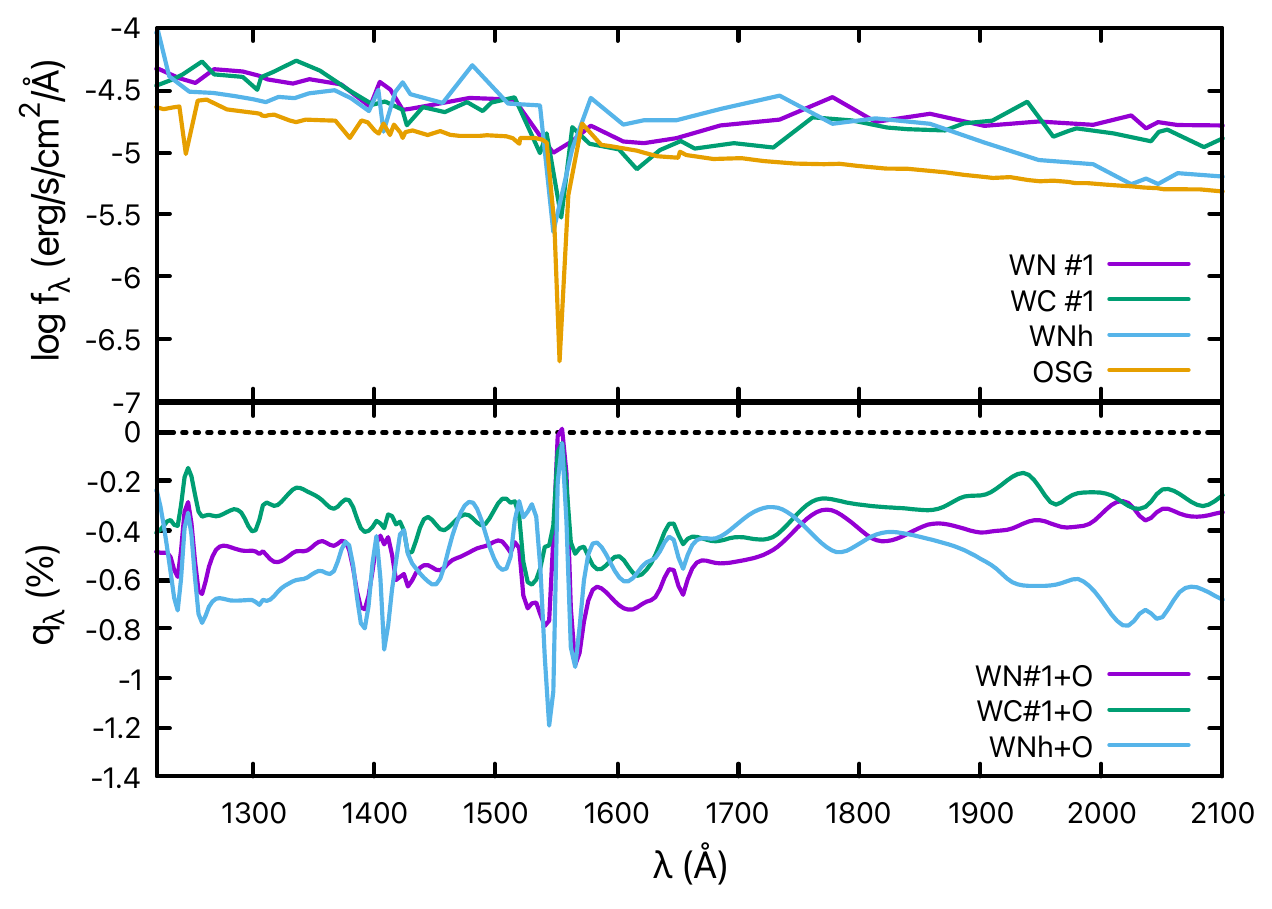}
\end{adjustwidth}
\caption{As in Fig.~\ref{fig2a}, now plotted only over the planned waveband for {\em Polstar}, from 1220-2100~\AA.
\label{fig2b}}
\end{figure}   
\unskip

\subsection{Circumstellar Electron Scattering}

The atmospheres and envelopes of hot stars allow for high ionization and scattering polarization by free electrons.  The capacity of a medium to produce a net polarization is a function of source asymmetry.  The amplitude of the polarization depends on both the degree of asymmetry and the optical depth.  Electron scattering has a small cross-section, and so continuum polarization arising from circumstellar envelopes is typically small, although there are exceptions such as the dense winds of Wolf-Rayet (WR) stars or the dense disks of Be~stars. 

An insighful result for circumstellar polarization was derived by \cite{1977A&A....57..141B} for any axisymmetric and optically-thin electron scattering circumstellar configuration.  The amount of polarization is given by

\begin{equation}
    p = \bar{\tau}\,\left(1-3\gamma\right)\,\sin^2 i,
    \label{eq:BM}
\end{equation}

\noindent where $\bar{\tau}$ is an directionally averaged optical depth of the axisymmetric envelope, $\gamma$ is a shape factor ranging from 0 to 1, with $1/3$ signifying a spherical envelope, and $i$ is the viewing inclination.  This expression shows a decoupling of the primary parameters that determine the polarization:  the amount of material, the deviation from spherical, and a perspective factor (namely, that a pole-on view for axisymmetry is a centro-symmetric intensity distribution so that polarization identically cancels).  The decoupling indicates significant degeneracy when seeking to interpret a polarization measurement.  While intrinsic polarization is an incontrovertible indicator of aspherical geometry, the power of polarization for discerning specifics of geometry becomes more possible with (a) temporal variability, (b) chromatic effects, (c) in combination with other diagnostics such as photometric variability, or (d) combinations of these.

There have been various expansions on the important work of \cite{1977A&A....57..141B}.  For example, the above expression assumes a point star of illumination; \cite{1987ApJ...317..290C} incorporated a correction for finite stellar size.  \cite{1999A&A...347..919A} explored polarizations resulting from stars with anisotropic illumination.  And \cite{1994A&A...289..492H, 1996A&A...308..521H, 1996ApJ...461..828W} investigated the effects of multiple scattering for thick envelopes in continuum and lines.

Another feature of equation~(\ref{eq:BM}) is that the predicted polarization is achromatic, because electron scattering is gray.  However, chromatic effects can arise for a variety of reasons, such as scattering envelopes that also have absorptive opacity \citep[e.g.,][]{1991ApJ...383L..67B}.  Another case is when there are multiple sources of illumination, such as binary systems in which the two stars have different spectral energy distributions (SEDs), as explored comprehensively by  \cite{1978A&A....68..415B} for thin scattering.  

As an example of interest for massive stars, \cite{2022ApJ...933....5I} considered the polarization from colliding winds binaries.  Using the solution of \cite{1996ApJ...469..729C} for an axisymmetric bow shock, \cite{2022ApJ...933....5I} was able to employ equation~(\ref{eq:BM}) as a special case of \cite{1978A&A....68..415B} and derive a formalism for computing polarized SEDs.  An example calculation appears in \cite{2022Ap&SS.367..118S}.  Figure~\ref{fig2a} and \ref{fig2b} provide new calculations in the form of O+WR colliding wind binaries.  Table~\ref{tab1} lists the properties of the stars for the several cases, including luminosity, mass-loss rates, terminal speeds, and stellar radii.  At binary separation $D$, parameters $p_{WR}$ and $p_O$ are the polarization contributions by each star to be combined, as described in \cite{2022ApJ...933....5I}.  

The synthetic spectra are based on PoWR models \citep{2012A&A...540A.144S, 2015A&A...579A..75T, 2019A&A...621A..85H} indicated by the model identifier in the Table as taken from the public online grid\footnote{www.astro.physik.uni-potsdam.de/PoWR/powrgrid1.php}.  Importantly, significant polarization only results for cases when the WR component has a high mass-loss rate, serving to elevate the optical depth for scattering. Models labeled WN\#2 and WC\#2 have small values of $p$ for both the O and the WR components even with a small orbital separation, so these cases are not shown in the Figures. For the other 3 cases of WN\#1, WC\#1, and WNh, the polarization is greatest at wavelengths where the O-star component is brightest, since the bowshock is spatially close to that component. These are the cases plotted with Figure~\ref{fig2a} from the EUV to the NIR, and Figure~\ref{fig2b} highlighting the passband planned for the FUV spectropolarimeter mission concept, {\em Polstar} (see \S~\ref{sec5}).  Note the figures are for a top-down view of the orbital plane at an orbital phase when $u_\lambda=0$.  

\begin{table}[t] 
\centering
\caption{Stellar and Orbital Properties for Binary Models \label{tab1}}
\begin{tabularx}{10.5cm}{lcccccc}
\toprule
\textbf{Property$^a$}	& \textbf{OSG}	& \textbf{WN \#1} & \textbf{WN \#2} & \textbf{WC \#1} & \textbf{WC \#2} & \textbf{WNh} \\
\midrule
Model & 40-42 & 04-12 & 17-12 & 06-14 & 17-14 & 13-21 \\
$\log L_\ast/L_\odot$ & 5.06 & 5.30 & 5.30 & 5.30 & 5.30 & 5.30 \\
$\log \dot{M}$ ($M_\odot/$yr) & $-7.0$ & $-4.23$ & $-6.18$ & $-4.34$ & $-5.99$ & $-4.44$  \\
$v_\infty$ (km/s) & 3158 & 1600 & 1600 & 2000 & 2000 & 1000 \\
$R_\ast$ ($R_\odot$) & 7.1 & 11.9 & 0.6 & 7.5 & 0.6 & 1.5 \\
$D$ ($R_\odot$) & --- & 250 & 30 & 250 & 30 & 250 \\
$p_{WR}$ (\%) & --- & +0.05 & $-0.01$ & +0.03 & $0.05$ & +0.09 \\
$p_{O}$ (\%) & --- & $-1.61$ & $-0.21 $ & $-1.04$ & $-0.24$ &  $-1.70$ \\
\bottomrule
\end{tabularx}

{\small $^a$ All the models are for Milky Way abundances, and the WNh star is\\ from the WNL-H50 group of PoWR models.}
\end{table}

\subsection{Continuum Polarization from Rapid Rotation}

Rapid rotation leads to a distortion of the star (oblateness) and a latitudinal flux distribution for the gravity darkening and polar brightening \citep{1924MNRAS..84..665V}.
Typically, the critical rotation of breakup is defined as $v_{\rm crit} = \sqrt{GM_\ast/R_\ast}$ at the equator of a star, for stellar mass $M_\ast$, stellar radius $R_\ast$, and gravitational constant $G$.  However, in approaching this limit, the oblateness of the star enlarges its equatorial radius, and gravity darkening can drive the equatorial region toward the Eddington limit, which also influences the definition of breakup rotation \citep[e.g.,][]{2000A&A...361..159M}.

Gravity darkening in particular presents challenges for inferring accurate values of $v \sin i$ from spectral lines.  While Be~stars are certainly rapid rotators, current measurements do not suggest near-critical rotations are typical \citep[e.g.,][]{2005ApJ...634..585C}.  However, gravity darkening can bias observational measures of $v \sin i$ values to underestimate the equatorial rotation speed \cite{2004MNRAS.350..189T}.  New approaches are needed to determine to what degree of criticality Be and Bn stars rotate.  It is already clear that some stars are quite close to critical \citep[e.g.,][]{2003A&A...407L..47D, 2017NatAs...1..690C, 2022ApJ...931...35S, 2024ApJ...972..103B, 2024MNRAS.529..374B}.

One way to directly measure the degree of critical rotation for hot stars is with continuum polarization.  With rotation increasingly near breakup, the distorted star along with redistribution of flux results in significant net polarizations from the stellar atmosphere.  While optical polarimetry has successfully inferred rapid rotation from continuum polarization at visible wavelengths in $\zeta$~Pup (O4I) and Regulus (B7V), the values are quite small \citep{2017NatAs...1..690C, 2024MNRAS.529..374B}.  By contrast the continuum polarization peaks toward FUV wavelengths \citep{1968ApJ...151.1051H,
1991ApJS...77..541C}.   An example of new calculations for B1V stars can be found in \cite{2024arXiv240915714I}.   As stars evolve toward lower surface gravities, the polarization will further increase \citep[e.g.,][]{2022MNRAS.513.1129L}.

\subsection{Polarization across Spectral Lines}

\begin{figure}[t]
\begin{adjustwidth}{-\extralength}{-3cm}
\centering\includegraphics[width=9 cm]{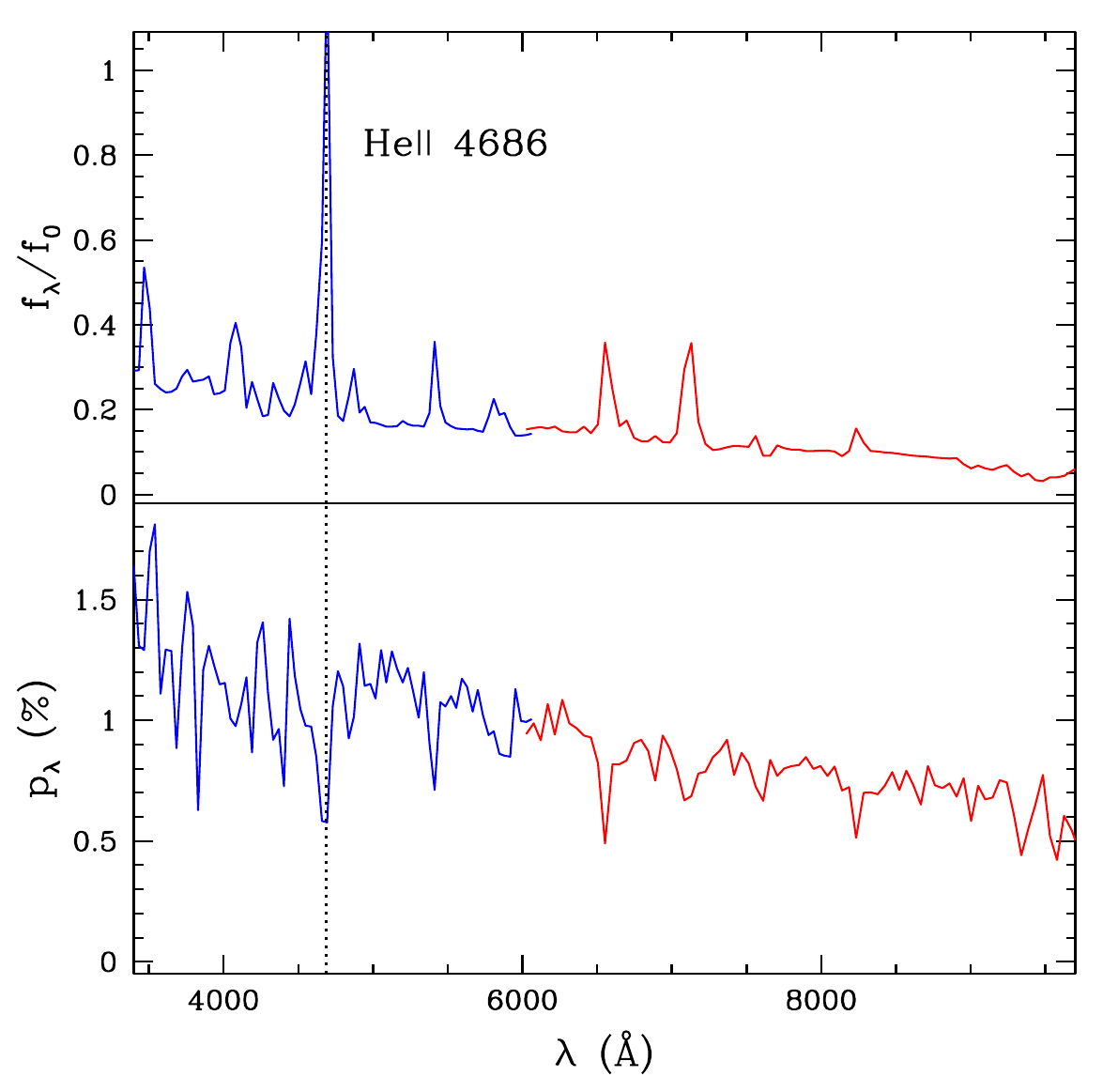}
\end{adjustwidth}
\caption{Illustration of the Line Effect from HPOL spectropolarimetry of the WNE star WR~134.  The upper panel shows the relative flux; the lower panel is the polarization.  Blue and red indicate the separate spectral gratings of HPOL \citep{2012AIPC.1429..226M}.  The vertical dotted line is for the He{\sc ii} 4686~{\AA} emission line, indicating a depression in polarization.
\label{fig3}}
\end{figure}   
\unskip

\subsubsection{The ``Line Effect''}

Spectral line formation in the winds and disks of massive stars can form over varying volumes.  Recombination lines in particular may form at considerable distance from the star.  However, electron polarization tends to form more local to the star, where density is higher.  As a result, line photons can escape the medium without being much scattered, as compared to the stellar continuum light.  Since polarization is expressed as a difference over the total, while the line photons do not contribute to the difference (i.e., line photons are unpolarized), they can increase the total used in the normalization.  As a result, polarization can be depressed across strong emission lines.  This dilution is also called the ``line effect'' \citep[e.g.,][]{1990ApJ...365L..19S}.

An example is shown in Figure~\ref{fig3} for the WN-type star WR~134 \citep{1992ApJ...387..347S}.  The data were taken with HPOL \citep{2012AIPC.1429..226M}.  The blue and red segments are for the shorter and longer wavelength gratings used by HPOL.  The upper panel is a normalized flux; lower is for the percent polarization.  The strong emission line of He{\sc ii} 4686~{\AA} is labeled with a vertical dotted line running across the two panels.  Note the line effect with diluted polarization in the lower panel.  Other emission lines (not labeled) also show this behavior.

The line effect is useful
for two reasons:  first it demonstrates the presence of intrinsic
stellar polarization, because ISP is not
subject to dilution. Second, the dilution
may be so severe as to yield an estimate for the ISP.  In terms of the ``categories'' of Figure~\ref{fig1}, a strong line effect can help determine the location of the crosses shown in the schematic Q-U diagrams of Figure~\ref{fig1}.  

\begin{figure}[t]
\begin{adjustwidth}{-\extralength}{-2cm}
\centering\includegraphics[width=8.5 cm]{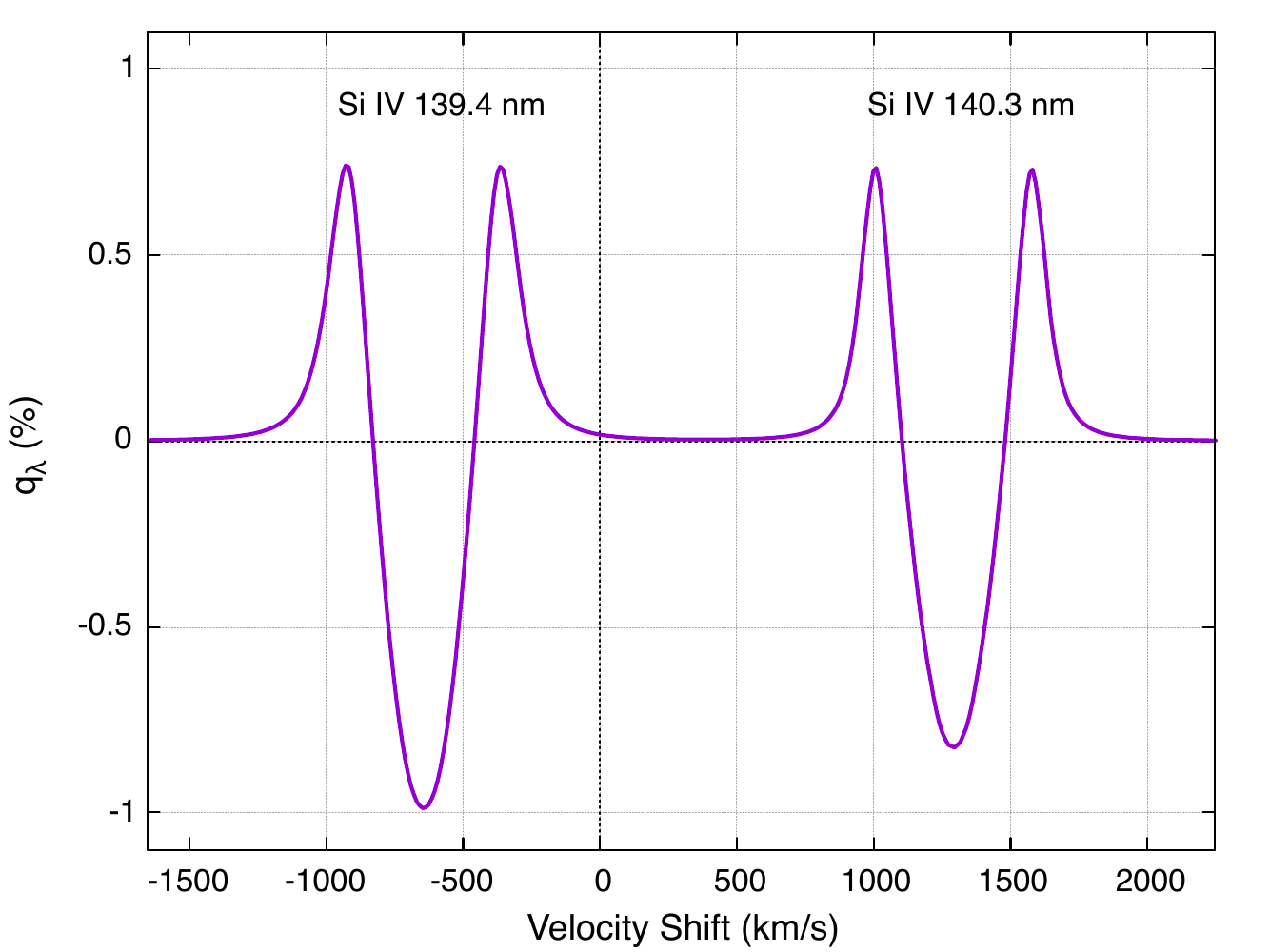}
\end{adjustwidth}
\caption{Illustration of the \"{O}hman Effect for the Si{\sc iv} 140~nm doublet.  The horizontal is velocity shift with zero being a weighted central wavelength for the doublet.  The polarization $q_\lambda$ is characteristically double-horned with a central peak of opposite sign than the horns.
This model is for $v_{\rm rot} = 360$~km/s viewing the star
edge-on (inclination $i=90^\circ$).
\label{fig4}}
\end{figure}   
\unskip

\subsubsection{The \"{O}hman Effect}

The \"{O}hman effect refers to a polarization change across rotationally
broadened photospheric lines \citep{1946ApJ...104..460O,
2024arXiv240711352B}.  The isovelocity zones of a rotating star are vertical strips, ranging from $-v\sin i$ to $+v\sin i$ and oriented parallel to the projected spin axis of the
star. A spectral resolution bin in velocity of $\Delta v$ maps to a strip of corresponding spatial width on the stellar atmosphere. (This is an approximate description, since thermal and turbulent broadening means strips will not have geometrically sharp edges.)

The Stokes~Q polarization displays a three-peak pattern, with two positive polarizations for the line wings, and one of negative sign at
line center, as illustrated in Figure~\ref{fig4}.
By contrast a Stokes~U polarization can result that is antisymmetric and zero at line center.  However, the U-polarization is weaker by an order of magnitude compared with $Q$ \citep[c.f., model predictions for $\epsilon$~Sgr in Fig.~12 of][]{2024ApJ...972..103B}.  The detailed profile shapes are sensitive
to the limb polarization profile, from center to limb, so can be
used to probe stellar atmosphere models as a function of rotation rate.

Polarization from the \"{O}hman effect is modest and can change fairly rapidly across the line profile.  The amplitude is much lower for optical lines than in the UV \citep{1991MNRAS.253..167C}.  At rotation speeds for which the star remains spherically symmetric, the net polarization in the line flux will cancel owing to symmetry, so use of the \"{O}hman diagnostic requires the line to be spectrally resolved. Note that the example in Figure~\ref{fig4} is for an edge-on inclination; generally the FWHM of the polarized profile is a good indicator of $v \sin i$.  

\subsubsection{The Hanle Effect}

The Hanle effect refers to a weak-field limit in which the Zeeman broadening has not fully lifted the degeneracy of the magnetic
sublevels, when the Zeeman splitting is only
of order the natural line broadening \citep{2001ASPC..236..161T}.  The Hanle effect operates in resonance scattering
lines and leads to linear polarization effects.  In classical terms, the harmonic oscillations of a radiating dipole is caused to precess at the Larmor frequency, $\omega_L$.  The precession leads to redistribution of scattered light and polarization that depends on the strength of the field, and its orientation with respect to the scattering angle.
Magnetic sensitivity is achieved typically when $\omega_L \sim A$, which depends on the $A$-value and tends to range from 0.1--100~G for UV lines relevant to massive stars.

However, the Hanle effect has yet to be observed in massive stars, although
it has been used for spectropolarimetic studies of the Sun
\citep[e.g.,][]{2013A&ARv..21...66S}.  There are three
key points for use of the Hanle effect
\citep{1997ApJ...486..550I}:  (1) The effect operates in resonance lines regardless of whether they are photospheric \citep{2012ApJ...760....7M} or circumstellar \citep{1997ApJ...486..550I}.  If circumstellar, the Hanle
effect acts as a direct probe of the magnetic field strength and topology at levels that are difficult for the Zeeman effect.  (2)
Resonance lines have vastly higher cross-sections than electron
scattering.  For targets with circumstellar media of low optical depth with electrons, resonance line polarization could still be significant.  (3) A multiline approach is key for extracting
information about the magnetic field.  For example, not all lines are Hanle-sensitive.  Several strong UV resonance lines for massive stars are Li-like doublets, for which one component produces no resonance polarization, and thus no Hanle effect.  That component serves as a control to infer magnetic
detection using the Hanle-sensitive one.  Multiple lines of different Hanle sensitivities can be used to map the magnetic field topology in respective regions of line formation.

\section{A Case Study Involving the Hanle Effect:  Magnetospheric Wind Channeling} \label{sec3}

Somewhat less than 10\% of nondegenerate massive stars possess significant magnetic fields, generally in excess of 100~G up to several kG \citep{2022Ap&SS.367..125F}.  The fields are consistent with a fossil origin, as opposed to dynamo activity that operates in the Sun and low-mass stars \citep{2015IAUS..305...61N}.  Numerous studies have explored the consequences of magnetism for these stars through MHD simulations or semi-analytic approaches \citep{2002ApJ...576..413U}.  Among the consequences are distorted wind flows, magnetospheric channeling, H$\alpha$ emission line production, non-thermal radio emissions, and rotational braking \citep{2013MNRAS.429..398P, 2009MNRAS.392.1022U, 2020MNRAS.499.5366O, 2022MNRAS.513.1449O, 2021MNRAS.506.5373E}.

As an example to illustrate use of the Hanle effect, we consider the magnetospheric channeling of a massive star wind \citep{2005MNRAS.357..251T, 2009MNRAS.392.1022U, 2012MNRAS.423L..21S, 2016MNRAS.462.3830O, 2020MNRAS.499.5366O}. Strong magnetic fields can direct and confine stellar wind flow, forcing material to corotate with the star.  The Hanle effect is a new diagnostic that could directly probe the field topology and flow kinematics beyond the stellar atmosphere.  While H$\alpha$ emission lines test models for Centrifugal Magnetospheres (CM), UV resonance lines can trace the pre-shocked magnetically channeled flow as well as the infalling material arising from the Dynamical Magnetospheric (DM) behavior \citep[e.g.,  see][for details about CMs and DMs]{2013MNRAS.429..398P}.  

\begin{table}[t] 
\centering
\caption{{\em Polstar} Targets for the Hanle Effect \label{tab2}}
\begin{tabularx}{7cm}{lcccc}
\toprule
\textbf{Name}  & \textbf{V-magn}	& \textbf{$B_{\rm p}$ (G)}	 & \textbf{$r_\infty/R_\ast$} & \textbf{$r_H/R_\ast$} \\
\midrule
$\tau$ Sco & 2.8 & 200 & 2.2 & 2.1 \\
$\beta$ Cep & 3.2 & 360 & 6.3 & 2.5 \\
$\beta$ Cen & 0.6 & 300 & 4.3 & 2.4 \\
16 Peg & 5.1 & $>500$ & 9.0 & 2.1 \\
\bottomrule
\end{tabularx}
\end{table}

\begin{figure}[t]
\begin{adjustwidth}{-\extralength}{-3cm}
\centering\includegraphics[width=8. cm]{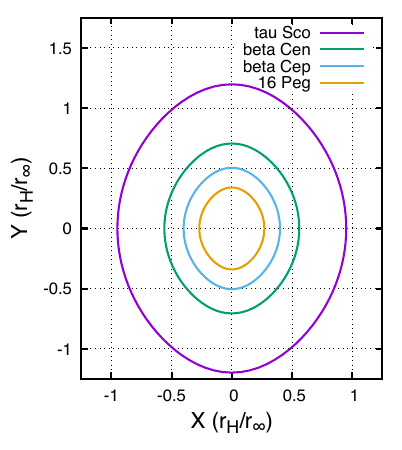}
\end{adjustwidth}
\caption{For the parameters in Table~\ref{tab1}, colored contours trace the location where $r_H=r_\infty$ as described in the text.  These curves are for the blue component of the C{\sc iv} doublet at 150~nm.  Adopted stellar parameters are given in Tab.~\ref{tab1}.
\label{fig5a}}
\end{figure}   
\unskip

\cite{2022Ap&SS.367..125F} provides a listing of magnetic field sensitivities for a variety of common scattering lines at UV wavelengths (see their Tab.~2).  That paper also provides an example calculation for a doublet P~Cygni line with associated polarization including the Hanle effect.  Here we seek to illustrate the spatial regions that would be probed by the Hanle effect for a selection of UV-bright stars using the C{\sc iv} 150~nm doublet as an example.

In the semi-classical prescription employing Larmor precession, the basic properties of scattered light can be described through a ``mixing angle'', as defined in \cite{2022Ap&SS.367..125F} to be

\begin{equation}
\tan \alpha_2 = \frac{B(\vec{r})}{B_H},
\end{equation}

\noindent where $B$ is the magnetic field and $B_H$ is the field sensitivity as given in Table~2 of \cite{2022Ap&SS.367..125F}.  For a dipole magnetic field, we introduce a corresponding radius $r_H$ where $B/B_H=1$ (i.e., $\alpha_2 = 45^\circ$), as given by

\begin{equation}
\frac{r_H}{R_\ast} = \left(\frac{B_{\rm p}}{B_H}\right)^{1/3}\,\left[ \frac{1+3\mu^2}{4}\right]^{1/6}, 
\label{eq:rH}
\end{equation}

\noindent where $B_{\rm p}$ is the strength of the polar magnetic field at the star, and $\mu = \cos \theta$ is the colatitude with respect to the star's magnetic axis.

In relation to the extent of the magnetosphere, it is useful to compare $r_H$ to the distance where the Alfven and wind terminal speeds are approximately equal, or $v_A \approx v_\infty$.  We introduce $r_\infty$ for this distance, as given by

\begin{equation}
r_\infty/R_\ast = \eta_\ast^{1/3}\,\left[ \frac{1+3\mu^2}{4}\right]^{1/3},
\end{equation} 

\noindent where $\eta_\ast$ is a parameter for how strong the field is compared to the wind as introduced by \cite{2002ApJ...576..413U}:

\begin{equation}
\eta_\ast = \frac{B^2_{\rm p}\,R_\ast^2}{\dot{M}\,v_\infty}.
\end{equation}

Using equation~(\ref{eq:rH}), Figure~\ref{fig5a} shows contours where $r_H=r_\infty$ for 4 UV-bright stars with modest surface magnetic fields, as labeled. Stellar and wind properties of the stars are given in Table~\ref{tab2}.  The $X-Y$ coordinates are defined so that the magnetic axis is a vertical line at $X=0$ and the magnetic equator is $Y=0$.  The contours are for the blue component of the C{\sc iv} doublet with $B_H=23$~G.  In this scaled diagram, it can be seen that this line will probe different regimes of the flow within the magnetosphere.  In the case of just one star, a selection of Hanle sensitive lines would produce a family of similar shaped contours to the different values in $B_H$. 

\section{Interstellar Polarization} \label{sec4}

As starlight passes through the interstellar medium (ISM), it
obtains a linear polarization, even if the star is spherical.
This arises from grain alignment by interstellar magnetism.  The grains then act like an absorptive filter that leaves a residual polarization orthogonal to the direction of grain alignment \citep{2015ARA&A..53..501A}.  The ISM thus imposes a polarization of some with an amplitude that is wavelength-dependent yet at fixed PA.
The form of this polarization is well-modeled by the Serkowski Law \citep{1975ApJ...196..261S}:

\begin{equation}
p_\lambda = p_{\rm max}\,e^{-K\,\ln^2(\lambda_{\rm max}/{\lambda})},
\end{equation}

\noindent where $p_{\rm max}$ is the maximum polarization at $\lambda=\lambda_{\rm max}$, and $K$ controls the width of the curve.  For a reasonably large sample of stars, $K \approx 1.66\, \lambda_{\rm max}(\mu m)$ with $\lambda_{\rm max}$ ranging from 0.4 to 0.8~$\mu m$ \citep[e.g.,][]{1992ApJ...386..562W}.

\begin{figure}[t]
\begin{adjustwidth}{-\extralength}{-2cm}
\centering\includegraphics[width=11. cm]{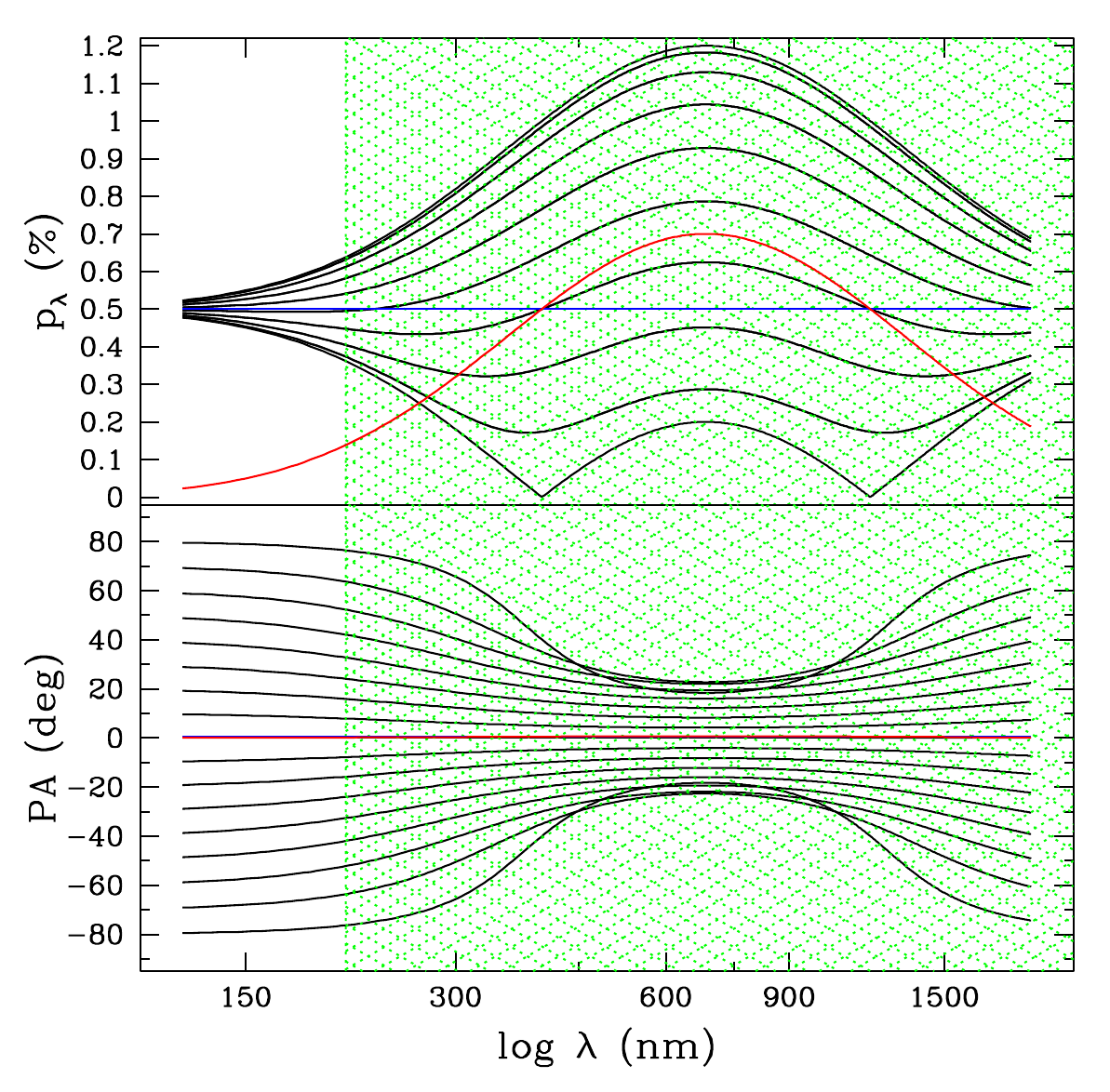}
\end{adjustwidth}
\caption{Illustration of how stellar and interstellar polarization combine.  The {\em Polstar} waveband is 122-210~nm.  The green shading is longward of 210~nm.
The dark blue line is electron scattering polarization at $p_\ast=0.5\%$; the red line is a Serkowski Law with $p_{\rm max}=0.7\%$, $K=1.15$, and $\lambda_{\rm max}=685$~nm.  The black curves represent the combined polarization for different values of stellar polarization position angle $\psi_\ast$, at $10^\circ$ increments, while $\psi_I=0^\circ$ is held fixed.  The upper panel is for polarization and the lower panel is for PA.  The lower panel highlights especially the transition of PAs tending toward the ISP in the visible versus toward the stellar in the FUV.
\label{fig5}}
\end{figure}   
\unskip

Figure~\ref{fig5} provides an heuristic example for combining the Serkowski Law with instrinic polarization from a star.  The upper panel is for polarization, and the lower is for PA (in degrees). The green shading signifies wavelengths that are longward of the FUV waveband planned for {\em Polstar} (see Sect.~\ref{sec5}).
The dark blue curve is a hypothetical intrinsic stellar polarization chosen at $p_\ast=0.5\%$ and assumed flat for electron scattering (i.e., no absorptive opacities).  The red curve is a Serkowski Law with parameters given in the caption, and at fixed PA of $\psi_I=0^\circ$.  The different black curves are for different PA values for the star, $\psi_\ast$, which result from the following expressions:

\begin{eqnarray}
q_\lambda & = & q_\ast + q_I = p_\ast\,\cos 2\psi_\ast + p_I, \\
u_\lambda & = & u_\ast + u_I =  p_\ast\,\sin 2\psi_\ast, \\
p_\lambda & = & \sqrt{q_\lambda^2 + u_\lambda^2}. \\
\tan 2 PA & = & u_\lambda/q_\lambda.
\end{eqnarray}

\noindent This exercise highlights the theme that ISP dominates at visible wavelengths and diminishes toward UV and IR wavelengths.  This is made clear from how the curves in P ``pinch'' toward $0^\circ$ at visible wavelengths, but show a range of values in the FUV as governed by the star.

\section{The {\em Polstar} UV Spectropolarimetry Mission Concept}
\label{sec5}

{\em Polstar} is a Small Explorer (SMEX) mission concept to study the role of rapid rotation and binary interaction among massive stars \citep{2024BSRSL..93..156I, 2024IAUS..361..633S, 2024sf2a.conf..457G}. The mission's focus will be on tracing angular momentum and mass exchange between stars, on its loss from the system, and on its transport within stars (either as ``capture'' for stars gaining angular momentum or as ``release'' for those losing it). {\em Polstar} will study massive stars at far ultraviolet (FUV) wavelengths with spectropolarimetry, combining spectroscopy for obtaining chemical abundances and systemic kinematics with linear polarimetry, which probes the unresolved geometry of stars and interacting binaries.

{\em Polstar's} polarimetric capability will cover a wavelength range of 122--210~nm, with spectroscopy obtained down to 117~nm. A rotating birefringent crystal assembly modulator will be combined with a Wollaston prism to obtain {\it s} and {\it p} polarizations. The light is dispersed to produce an Echelle spectrum at $R\approx 20,000$ allowing for velocity resolutions of 15 km/s.  While Stokes V will be obtained, the design will be optimized for obtaining Stokes I, Q, and U fluxes. With an aperture of 40~cm, the goal is to achieve a precision of 0.0003 in fractional polarization.

The FUV waveband is chosen for several reasons. First, the FUV contains numerous high-opacity resonance lines, which are commonly observed among massive stars, and which are suitable for tracing the kinematics of winds, disks, and accretion streams for individual and binary stellar systems.  The resonance scattering lines are also suitable for use of the Hanle effect, which is not the case for lines found in the optical regime.  Likewise, the polarization for the \"{O}hman effect is greatest at FUV wavelengths.  Importantly, the effects of rapid to critical rotation have steeply rising and strong continuum polarization (of order 1\% or more in some cases) toward FUV wavelengths.  The FUV is also ideal for identifying hot subdwarf OB (sdO/sdB) companions to Be and Bn stars.

It is useful to compare the {\em Polstar} concept with the past missions of the Wisconsin Ultraviolet Photopolarimetry Experiment \citep[{\em WUPPE} ][]{1994SPIE.2010....2N} and the International Ultraviolet Explorer \citep[{\em IUE}][]{1978Natur.275..372B}.  {\em WUPPE} performed low-resolution linear polarimetry at $R\sim 1000$, largely as a broad ``snapshot'' survey. It observed a range of spectral classes, AGN, and conducted a detailed study of the ISP at UV wavelengths. In contrast, the strategy for {\em Polstar} will also include monitoring for chosen stars over rotational and orbital timescales. {\em Polstar's} high spectral resolution will for the first time allow the combination of both geometric and kinematic probes across spectral lines at FUV wavelengths.

{\em IUE} was a long running and highly productive UV spectroscopic mission.  However, it suffered from fixed pattern noise that capped the SNR to $\sim 35$ for any given exposure.  The popularity of {\em IUE} also limited its capability for conducting long-term ``staring'' mode or phase monitoring observations of massive stars, with a significant exception being \cite{1995ApJ...452L..53M}.  {\em Polstar} will foremost be capable of binning and stacking datasets to achieve incredibly high SNR approaching $10^4$ for the brightest sources.  This will afford both stunning dynamical spectra along with unprecedented polarization measurements at FUV wavelengths.

Led by Lockheed Martin in strategic partnership with the Centre national d'{\'e}tudes spatiales (CNES), together with a science team of international experts on massive stars, UV spectroscopy, and polarimetry, the plan for {\em Polstar} will involve high spectral resolution, high cadence, and dense phase sampling of individual and interacting binary systems, with a significant focus on the B spectral class.  The mission will answer the questions: How close to criticality can massive stars rotate?  What is the frequency with which the rapidly rotating Bn and Be stars have stripped core companions?  What governs the efficiency of angular momentum exchange in massive interacting binaries?  Additionally, the mission concept anticipates a guest observer program, providing the astronomical community with a new and unique device for the study of unresolved astrophysical sources. 




\dataavailability{No new data were involved in this study. }

\acknowledgments{
C.E. gratefully acknowledges support for this work provided by NASA through grant number HST-AR-15794.005, from the Space Telescope Science Institute, which is operated by AURA, Inc., under NASA contract NAS 5-26555.}

\conflictsofinterest{``All of the authors are currently members of the science team for the {\em Polstar} mission concept.''} 

\begin{adjustwidth}{-\extralength}{0cm}

\reftitle{References}


\bibliography{almaty}

\begin{thebibliography}{999}

\bibitem[{Langer}(2012)]{2012ARA&A..50..107L}
{Langer}, N.
\newblock {Presupernova Evolution of Massive Single and Binary Stars}.
\newblock {\em ARA\&A} {\bf 2012}, {\em 50},~107--164,
  \href{http://arxiv.org/abs/1206.5443}{{\normalfont
  [arXiv:astro-ph.SR/1206.5443]}}.
\newblock {\url{https://doi.org/10.1146/annurev-astro-081811-125534}}.

\bibitem[{Nomoto} et~al.(2013){Nomoto}, {Kobayashi}, and
  {Tominaga}]{2013ARA&A..51..457N}
{Nomoto}, K.; {Kobayashi}, C.; {Tominaga}, N.
\newblock {Nucleosynthesis in Stars and the Chemical Enrichment of Galaxies}.
\newblock {\em ARA\&A} {\bf 2013}, {\em 51},~457--509.
\newblock {\url{https://doi.org/10.1146/annurev-astro-082812-140956}}.

\bibitem[{Belczynski} et~al.(2010){Belczynski}, {Bulik}, {Fryer}, {Ruiter},
  {Valsecchi}, {Vink}, and {Hurley}]{2010ApJ...714.1217B}
{Belczynski}, K.; {Bulik}, T.; {Fryer}, C.L.; {Ruiter}, A.; {Valsecchi}, F.;
  {Vink}, J.S.; {Hurley}, J.R.
\newblock {On the Maximum Mass of Stellar Black Holes}.
\newblock {\em ApJ} {\bf 2010}, {\em 714},~1217--1226,
  \href{http://arxiv.org/abs/0904.2784}{{\normalfont
  [arXiv:astro-ph.SR/0904.2784]}}.
\newblock {\url{https://doi.org/10.1088/0004-637X/714/2/1217}}.

\bibitem[{Vink}(2022)]{2022ARA&A..60..203V}
{Vink}, J.S.
\newblock {Theory and Diagnostics of Hot Star Mass Loss}.
\newblock {\em ARA\&A} {\bf 2022}, {\em 60},~203--246,
  \href{http://arxiv.org/abs/2109.08164}{{\normalfont
  [arXiv:astro-ph.SR/2109.08164]}}.
\newblock {\url{https://doi.org/10.1146/annurev-astro-052920-094949}}.

\bibitem[{Sana} et~al.(2012){Sana}, {de Mink}, {de Koter}, {Langer}, {Evans},
  {Gieles}, {Gosset}, {Izzard}, {Le Bouquin}, and
  {Schneider}]{2012Sci...337..444S}
{Sana}, H.; {de Mink}, S.E.; {de Koter}, A.; {Langer}, N.; {Evans}, C.J.;
  {Gieles}, M.; {Gosset}, E.; {Izzard}, R.G.; {Le Bouquin}, J.B.; {Schneider},
  F.R.N.
\newblock {Binary Interaction Dominates the Evolution of Massive Stars}.
\newblock {\em Science} {\bf 2012}, {\em 337},~444,
  \href{http://arxiv.org/abs/1207.6397}{{\normalfont
  [arXiv:astro-ph.SR/1207.6397]}}.
\newblock {\url{https://doi.org/10.1126/science.1223344}}.

\bibitem[{Kummer} et~al.(2023){Kummer}, {Toonen}, and {de
  Koter}]{2023A&A...678A..60K}
{Kummer}, F.; {Toonen}, S.; {de Koter}, A.
\newblock {The main evolutionary pathways of massive hierarchical triple
  stars}.
\newblock {\em A\&A} {\bf 2023}, {\em 678},~A60,
  \href{http://arxiv.org/abs/2306.09400}{{\normalfont
  [arXiv:astro-ph.SR/2306.09400]}}.
\newblock {\url{https://doi.org/10.1051/0004-6361/202347179}}.

\bibitem[{Ud-Doula} et~al.(2009){Ud-Doula}, {Owocki}, and
  {Townsend}]{2009MNRAS.392.1022U}
{Ud-Doula}, A.; {Owocki}, S.P.; {Townsend}, R.H.D.
\newblock {Dynamical simulations of magnetically channelled line-driven stellar
  winds - III. Angular momentum loss and rotational spin-down}.
\newblock {\em MNRAS} {\bf 2009}, {\em 392},~1022--1033,
  \href{http://arxiv.org/abs/0810.4247}{{\normalfont
  [arXiv:astro-ph/0810.4247]}}.
\newblock {\url{https://doi.org/10.1111/j.1365-2966.2008.14134.x}}.

\bibitem[{Renzo} et~al.(2019){Renzo}, {Zapartas}, {de Mink}, {G{\"o}tberg},
  {Justham}, {Farmer}, {Izzard}, {Toonen}, and {Sana}]{2019A&A...624A..66R}
{Renzo}, M.; {Zapartas}, E.; {de Mink}, S.E.; {G{\"o}tberg}, Y.; {Justham}, S.;
  {Farmer}, R.J.; {Izzard}, R.G.; {Toonen}, S.; {Sana}, H.
\newblock {Massive runaway and walkaway stars. A study of the kinematical
  imprints of the physical processes governing the evolution and explosion of
  their binary progenitors}.
\newblock {\em A\&A} {\bf 2019}, {\em 624},~A66,
  \href{http://arxiv.org/abs/1804.09164}{{\normalfont
  [arXiv:astro-ph.SR/1804.09164]}}.
\newblock {\url{https://doi.org/10.1051/0004-6361/201833297}}.

\bibitem[{Penny}(1996)]{1996ApJ...463..737P}
{Penny}, L.R.
\newblock {Projected Rotational Velocities of O-Type Stars}.
\newblock {\em ApJ} {\bf 1996}, {\em 463},~737.
\newblock {\url{https://doi.org/10.1086/177286}}.

\bibitem[{Huang} et~al.(2010){Huang}, {Gies}, and
  {McSwain}]{2010ApJ...722..605H}
{Huang}, W.; {Gies}, D.R.; {McSwain}, M.V.
\newblock {A Stellar Rotation Census of B Stars: From ZAMS to TAMS}.
\newblock {\em ApJ} {\bf 2010}, {\em 722},~605--619,
  \href{http://arxiv.org/abs/1008.1761}{{\normalfont
  [arXiv:astro-ph.SR/1008.1761]}}.
\newblock {\url{https://doi.org/10.1088/0004-637X/722/1/605}}.

\bibitem[{Nathaniel} et~al.(2024){Nathaniel}, {Vigna-G{\'o}mez}, {Grichener},
  {Farmer}, {Renzo}, and {Everson}]{2024arXiv240711680N}
{Nathaniel}, K.; {Vigna-G{\'o}mez}, A.; {Grichener}, A.; {Farmer}, R.; {Renzo},
  M.; {Everson}, R.W.
\newblock {Population synthesis of Thorne-{\.Z}ytkow objects: Rejuvenated
  donors and unexplored progenitors in the common envelope formation channel}.
\newblock {\em arXiv e-prints} {\bf 2024}, p. arXiv:2407.11680,
  \href{http://arxiv.org/abs/2407.11680}{{\normalfont
  [arXiv:astro-ph.SR/2407.11680]}}.
\newblock {\url{https://doi.org/10.48550/arXiv.2407.11680}}.

\bibitem[{Podsiadlowski} et~al.(1992){Podsiadlowski}, {Joss}, and
  {Hsu}]{1992ApJ...391..246P}
{Podsiadlowski}, P.; {Joss}, P.C.; {Hsu}, J.J.L.
\newblock {Presupernova Evolution in Massive Interacting Binaries}.
\newblock {\em ApJ} {\bf 1992}, {\em 391},~246.
\newblock {\url{https://doi.org/10.1086/171341}}.

\bibitem[{Toonen} et~al.(2020){Toonen}, {Portegies Zwart}, {Hamers}, and
  {Bandopadhyay}]{2020A&A...640A..16T}
{Toonen}, S.; {Portegies Zwart}, S.; {Hamers}, A.S.; {Bandopadhyay}, D.
\newblock {The evolution of stellar triples. The most common evolutionary
  pathways}.
\newblock {\em A\&A} {\bf 2020}, {\em 640},~A16,
  \href{http://arxiv.org/abs/2004.07848}{{\normalfont
  [arXiv:astro-ph.SR/2004.07848]}}.
\newblock {\url{https://doi.org/10.1051/0004-6361/201936835}}.

\bibitem[{Wade} et~al.(2016){Wade}, {Neiner}, {Alecian}, {Grunhut}, {Petit},
  {Batz}, {Bohlender}, {Cohen}, {Henrichs}, {Kochukhov}, {Landstreet},
  {Manset}, {Martins}, {Mathis}, {Oksala}, {Owocki}, {Rivinius}, {Shultz},
  {Sundqvist}, {Townsend}, {ud-Doula}, {Bouret}, {Braithwaite}, {Briquet},
  {Carciofi}, {David-Uraz}, {Folsom}, {Fullerton}, {Leroy}, {Marcolino},
  {Moffat}, {Naz{\'e}}, {Louis}, {Auri{\`e}re}, {Bagnulo}, {Bailey},
  {Barb{\'a}}, {Blaz{\`e}re}, {B{\"o}hm}, {Catala}, {Donati}, {Ferrario},
  {Harrington}, {Howarth}, {Ignace}, {Kaper}, {L{\"u}ftinger}, {Prinja},
  {Vink}, {Weiss}, and {Yakunin}]{2016MNRAS.456....2W}
{Wade}, G.A.; {Neiner}, C.; {Alecian}, E.; {Grunhut}, J.H.; {Petit}, V.;
  {Batz}, B.d.; {Bohlender}, D.A.; {Cohen}, D.H.; {Henrichs}, H.F.;
  {Kochukhov}, O.;  et~al.
\newblock {The MiMeS survey of magnetism in massive stars: introduction and
  overview}.
\newblock {\em MNRAS} {\bf 2016}, {\em 456},~2--22,
  \href{http://arxiv.org/abs/1511.08425}{{\normalfont
  [arXiv:astro-ph.SR/1511.08425]}}.
\newblock {\url{https://doi.org/10.1093/mnras/stv2568}}.

\bibitem[{Sch{\"o}ller} et~al.(2017){Sch{\"o}ller}, {Hubrig}, {Fossati},
  {Carroll}, {Briquet}, {Oskinova}, {J{\"a}rvinen}, {Ilyin}, {Castro}, {Morel},
  {Langer}, {Przybilla}, {Nieva}, {Kholtygin}, {Sana}, {Herrero}, {Barb{\'a}},
  {de Koter}, and {BOB Collaboration}]{2017A&A...599A..66S}
{Sch{\"o}ller}, M.; {Hubrig}, S.; {Fossati}, L.; {Carroll}, T.A.; {Briquet},
  M.; {Oskinova}, L.M.; {J{\"a}rvinen}, S.; {Ilyin}, I.; {Castro}, N.; {Morel},
  T.;  et~al.
\newblock {B fields in OB stars (BOB): Concluding the FORS 2 observing
  campaign}.
\newblock {\em \aap} {\bf 2017}, {\em 599},~A66,
  \href{http://arxiv.org/abs/1611.04502}{{\normalfont
  [arXiv:astro-ph.SR/1611.04502]}}.
\newblock {\url{https://doi.org/10.1051/0004-6361/201628905}}.

\bibitem[{Chandrasekhar}(1960)]{1960ratr.book.....C}
{Chandrasekhar}, S.
\newblock {\em {Radiative transfer}};  1960.

\bibitem[{Brown} and {McLean}(1977)]{1977A&A....57..141B}
{Brown}, J.C.; {McLean}, I.S.
\newblock {Polarisation by Thomson Scattering in Optically Thin Stellar
  Envelopes. I. Source Star at Centre of Axisymmetric Envelope}.
\newblock {\em \aap} {\bf 1977}, {\em 57},~141.

\bibitem[{Cassinelli} et~al.(1987){Cassinelli}, {Nordsieck}, and
  {Murison}]{1987ApJ...317..290C}
{Cassinelli}, J.P.; {Nordsieck}, K.H.; {Murison}, M.A.
\newblock {Polarization of Light Scattered from the Winds of Early-Type Stars}.
\newblock {\em \apj} {\bf 1987}, {\em 317},~290.
\newblock {\url{https://doi.org/10.1086/165277}}.

\bibitem[{Al-Malki} et~al.(1999){Al-Malki}, {Simmons}, {Ignace}, {Brown}, and
  {Clarke}]{1999A&A...347..919A}
{Al-Malki}, M.B.; {Simmons}, J.F.L.; {Ignace}, R.; {Brown}, J.C.; {Clarke}, D.
\newblock {Scattering polarization due to light source anisotropy. I. Large
  spherical envelope}.
\newblock {\em \aap} {\bf 1999}, {\em 347},~919--926.

\bibitem[{Hillier}(1994)]{1994A&A...289..492H}
{Hillier}, D.J.
\newblock {The calculation of continuum polarization due to the Rayleigh
  scattering phase matrix in multi-scattering axisymmetric envelopes.}
\newblock {\em \aap} {\bf 1994}, {\em 289},~492--504.

\bibitem[{Hillier}(1996)]{1996A&A...308..521H}
{Hillier}, D.J.
\newblock {The calculation of line polarization due to scattering by electrons
  in multi-scattering axisymmetric envelopes.}
\newblock {\em \aap} {\bf 1996}, {\em 308},~521--534.

\bibitem[{Wood} et~al.(1996){Wood}, {Bjorkman}, {Whitney}, and
  {Code}]{1996ApJ...461..828W}
{Wood}, K.; {Bjorkman}, J.E.; {Whitney}, B.A.; {Code}, A.D.
\newblock {The Effect of Multiple Scattering on the Polarization from
  Axisymmetric Circumstellar Envelopes. I. Pure Thomson Scattering Envelopes}.
\newblock {\em \apj} {\bf 1996}, {\em 461},~828.
\newblock {\url{https://doi.org/10.1086/177105}}.

\bibitem[{Bjorkman} et~al.(1991){Bjorkman}, {Nordsieck}, {Code}, {Anderson},
  {Babler}, {Clayton}, {Magalhaes}, {Meade}, {Nook}, {Schulte-Ladbeck},
  {Taylor}, and {Whitney}]{1991ApJ...383L..67B}
{Bjorkman}, K.S.; {Nordsieck}, K.H.; {Code}, A.D.; {Anderson}, C.M.; {Babler},
  B.L.; {Clayton}, G.C.; {Magalhaes}, A.M.; {Meade}, M.R.; {Nook}, M.A.;
  {Schulte-Ladbeck}, R.E.;  et~al.
\newblock {First Ultraviolet Spectropolarimetry of Be Stars from the Wisconsin
  Ultraviolet Photo-Polarimeter Experiment}.
\newblock {\em \apjl} {\bf 1991}, {\em 383},~L67.
\newblock {\url{https://doi.org/10.1086/186243}}.

\bibitem[{Brown} et~al.(1978){Brown}, {McLean}, and
  {Emslie}]{1978A&A....68..415B}
{Brown}, J.C.; {McLean}, I.S.; {Emslie}, A.G.
\newblock {Polarisation by Thomson scattering in optically thin stellar
  envelopes. II. Binary and multiple star envelopes and the determination of
  binary inclinations.}
\newblock {\em \aap} {\bf 1978}, {\em 68},~415--427.

\bibitem[{Ignace} et~al.(2022){Ignace}, {Fullard}, {Shrestha}, {Naz{\'e}},
  {Gayley}, {Hoffman}, {Lomax}, and {St-Louis}]{2022ApJ...933....5I}
{Ignace}, R.; {Fullard}, A.; {Shrestha}, M.; {Naz{\'e}}, Y.; {Gayley}, K.;
  {Hoffman}, J.L.; {Lomax}, J.R.; {St-Louis}, N.
\newblock {Modeling the Optical to Ultraviolet Polarimetric Variability from
  Thomson Scattering in Colliding-wind Binaries}.
\newblock {\em \apj} {\bf 2022}, {\em 933},~5,
  \href{http://arxiv.org/abs/2205.07612}{{\normalfont
  [arXiv:astro-ph.SR/2205.07612]}}.
\newblock {\url{https://doi.org/10.3847/1538-4357/ac6fce}}.

\bibitem[{Cant{\'o}} et~al.(1996){Cant{\'o}}, {Raga}, and
  {Wilkin}]{1996ApJ...469..729C}
{Cant{\'o}}, J.; {Raga}, A.C.; {Wilkin}, F.P.
\newblock {Exact, Algebraic Solutions of the Thin-Shell Two-Wind Interaction
  Problem}.
\newblock {\em \apj} {\bf 1996}, {\em 469},~729.
\newblock {\url{https://doi.org/10.1086/177820}}.

\bibitem[{St-Louis} et~al.(2022){St-Louis}, {Gayley}, {Hillier}, {Ignace},
  {Jones}, {David-Uraz}, {Richardson}, {Vink}, {Peters}, {Hoffman}, {Naz{\'e}},
  {Stevance}, {Shenar}, {Fullard}, {Lomax}, and {Scowen}]{2022Ap&SS.367..118S}
{St-Louis}, N.; {Gayley}, K.; {Hillier}, D.J.; {Ignace}, R.; {Jones}, C.E.;
  {David-Uraz}, A.; {Richardson}, N.D.; {Vink}, J.S.; {Peters}, G.J.;
  {Hoffman}, J.L.;  et~al.
\newblock {UV spectropolarimetry with Polstar: massive star binary colliding
  winds}.
\newblock {\em \apss} {\bf 2022}, {\em 367},~118,
  \href{http://arxiv.org/abs/2207.07163}{{\normalfont
  [arXiv:astro-ph.SR/2207.07163]}}.
\newblock {\url{https://doi.org/10.1007/s10509-022-04102-0}}.

\bibitem[{Sander} et~al.(2012){Sander}, {Hamann}, and
  {Todt}]{2012A&A...540A.144S}
{Sander}, A.; {Hamann}, W.R.; {Todt}, H.
\newblock {The Galactic WC stars. Stellar parameters from spectral analyses
  indicate a new evolutionary sequence}.
\newblock {\em \aap} {\bf 2012}, {\em 540},~A144,
  \href{http://arxiv.org/abs/1201.6354}{{\normalfont
  [arXiv:astro-ph.SR/1201.6354]}}.
\newblock {\url{https://doi.org/10.1051/0004-6361/201117830}}.

\bibitem[{Todt} et~al.(2015){Todt}, {Sander}, {Hainich}, {Hamann}, {Quade}, and
  {Shenar}]{2015A&A...579A..75T}
{Todt}, H.; {Sander}, A.; {Hainich}, R.; {Hamann}, W.R.; {Quade}, M.; {Shenar},
  T.
\newblock {Potsdam Wolf-Rayet model atmosphere grids for WN stars}.
\newblock {\em \aap} {\bf 2015}, {\em 579},~A75.
\newblock {\url{https://doi.org/10.1051/0004-6361/201526253}}.

\bibitem[{Hainich} et~al.(2019){Hainich}, {Ramachandran}, {Shenar}, {Sander},
  {Todt}, {Gruner}, {Oskinova}, and {Hamann}]{2019A&A...621A..85H}
{Hainich}, R.; {Ramachandran}, V.; {Shenar}, T.; {Sander}, A.A.C.; {Todt}, H.;
  {Gruner}, D.; {Oskinova}, L.M.; {Hamann}, W.R.
\newblock {PoWR grids of non-LTE model atmospheres for OB-type stars of various
  metallicities}.
\newblock {\em \aap} {\bf 2019}, {\em 621},~A85,
  \href{http://arxiv.org/abs/1811.06307}{{\normalfont
  [arXiv:astro-ph.SR/1811.06307]}}.
\newblock {\url{https://doi.org/10.1051/0004-6361/201833787}}.

\bibitem[{von Zeipel}(1924)]{1924MNRAS..84..665V}
{von Zeipel}, H.
\newblock {The radiative equilibrium of a rotating system of gaseous masses}.
\newblock {\em MNRAS} {\bf 1924}, {\em 84},~665--683.
\newblock {\url{https://doi.org/10.1093/mnras/84.9.665}}.

\bibitem[{Maeder} and {Meynet}(2000)]{2000A&A...361..159M}
{Maeder}, A.; {Meynet}, G.
\newblock {Stellar evolution with rotation. VI. The Eddington and Omega
  -limits, the rotational mass loss for OB and LBV stars}.
\newblock {\em A\&A} {\bf 2000}, {\em 361},~159--166,
  \href{http://arxiv.org/abs/astro-ph/0006405}{{\normalfont
  [arXiv:astro-ph/astro-ph/0006405]}}.
\newblock {\url{https://doi.org/10.48550/arXiv.astro-ph/0006405}}.

\bibitem[{Cranmer}(2005)]{2005ApJ...634..585C}
{Cranmer}, S.R.
\newblock {A Statistical Study of Threshold Rotation Rates for the Formation of
  Disks around Be Stars}.
\newblock {\em \apj} {\bf 2005}, {\em 634},~585--601,
  \href{http://arxiv.org/abs/astro-ph/0507718}{{\normalfont
  [arXiv:astro-ph/astro-ph/0507718]}}.
\newblock {\url{https://doi.org/10.1086/491696}}.

\bibitem[{Townsend} et~al.(2004){Townsend}, {Owocki}, and
  {Howarth}]{2004MNRAS.350..189T}
{Townsend}, R.H.D.; {Owocki}, S.P.; {Howarth}, I.D.
\newblock {Be-star rotation: how close to critical?}
\newblock {\em \mnras} {\bf 2004}, {\em 350},~189--195,
  \href{http://arxiv.org/abs/astro-ph/0312113}{{\normalfont
  [arXiv:astro-ph/astro-ph/0312113]}}.
\newblock {\url{https://doi.org/10.1111/j.1365-2966.2004.07627.x}}.

\bibitem[{Domiciano de Souza} et~al.(2003){Domiciano de Souza}, {Kervella},
  {Jankov}, {Abe}, {Vakili}, {di Folco}, and {Paresce}]{2003A&A...407L..47D}
{Domiciano de Souza}, A.; {Kervella}, P.; {Jankov}, S.; {Abe}, L.; {Vakili},
  F.; {di Folco}, E.; {Paresce}, F.
\newblock {The spinning-top Be star Achernar from VLTI-VINCI}.
\newblock {\em A\&A} {\bf 2003}, {\em 407},~L47--L50,
  \href{http://arxiv.org/abs/astro-ph/0306277}{{\normalfont
  [arXiv:astro-ph/astro-ph/0306277]}}.
\newblock {\url{https://doi.org/10.1051/0004-6361:20030786}}.

\bibitem[{Cotton} et~al.(2017){Cotton}, {Bailey}, {Howarth}, {Bott},
  {Kedziora-Chudczer}, {Lucas}, and {Hough}]{2017NatAs...1..690C}
{Cotton}, D.V.; {Bailey}, J.; {Howarth}, I.D.; {Bott}, K.; {Kedziora-Chudczer},
  L.; {Lucas}, P.W.; {Hough}, J.H.
\newblock {Polarization due to rotational distortion in the bright star
  Regulus}.
\newblock {\em Nature Astronomy} {\bf 2017}, {\em 1},~690--696,
  \href{http://arxiv.org/abs/1804.06576}{{\normalfont
  [arXiv:astro-ph.SR/1804.06576]}}.
\newblock {\url{https://doi.org/10.1038/s41550-017-0238-6}}.

\bibitem[{Shepard} et~al.(2022){Shepard}, {Gies}, {Kaper}, and {De
  Koter}]{2022ApJ...931...35S}
{Shepard}, K.; {Gies}, D.R.; {Kaper}, L.; {De Koter}, A.
\newblock {Spectroscopic Line Modeling of the Fastest Rotating O-type Stars}.
\newblock {\em \apj} {\bf 2022}, {\em 931},~35,
  \href{http://arxiv.org/abs/2204.07473}{{\normalfont
  [arXiv:astro-ph.SR/2204.07473]}}.
\newblock {\url{https://doi.org/10.3847/1538-4357/ac66e6}}.

\bibitem[{Bailey} et~al.(2024{\natexlab{a}}){Bailey}, {Lewis}, {Howarth},
  {Cotton}, {Marshall}, and {Kedziora-Chudczer}]{2024ApJ...972..103B}
{Bailey}, J.; {Lewis}, F.; {Howarth}, I.D.; {Cotton}, D.V.; {Marshall}, J.P.;
  {Kedziora-Chudczer}, L.
\newblock {$\epsilon$ Sagittarii: An Extreme Rapid Rotator with a Decretion
  Disk}.
\newblock {\em \apj} {\bf 2024}, {\em 972},~103,
  \href{http://arxiv.org/abs/2407.11352}{{\normalfont
  [arXiv:astro-ph.SR/2407.11352]}}.
\newblock {\url{https://doi.org/10.3847/1538-4357/ad630b}}.

\bibitem[{Bailey} et~al.(2024{\natexlab{b}}){Bailey}, {Howarth}, {Cotton},
  {Kedziora-Chudczer}, {De Horta}, {Martell}, {Eldridge}, and
  {Luckas}]{2024MNRAS.529..374B}
{Bailey}, J.; {Howarth}, I.D.; {Cotton}, D.V.; {Kedziora-Chudczer}, L.; {De
  Horta}, A.; {Martell}, S.L.; {Eldridge}, C.; {Luckas}, P.
\newblock {Rapid polarization variations in the O4 supergiant
  {\ensuremath{\zeta}} Puppis}.
\newblock {\em MNRAS} {\bf 2024}, {\em 529},~374--392,
  \href{http://arxiv.org/abs/2402.13383}{{\normalfont
  [arXiv:astro-ph.SR/2402.13383]}}.
\newblock {\url{https://doi.org/10.1093/mnras/stae548}}.

\bibitem[{Harrington} and {Collins}(1968)]{1968ApJ...151.1051H}
{Harrington}, J.P.; {Collins}, George~W., I.
\newblock {Intrinsic Polarization of Rapidly Rotating Early-Type Stars}.
\newblock {\em ApJ} {\bf 1968}, {\em 151},~1051.
\newblock {\url{https://doi.org/10.1086/149504}}.

\bibitem[{Collins} et~al.(1991){Collins}, {Truax}, and
  {Cranmer}]{1991ApJS...77..541C}
{Collins}, George~W., I.; {Truax}, R.J.; {Cranmer}, S.R.
\newblock {Model Atmospheres for Rotating B Stars}.
\newblock {\em ApJS} {\bf 1991}, {\em 77},~541.
\newblock {\url{https://doi.org/10.1086/191616}}.

\bibitem[{Ignace} and {Scowen}(2024)]{2024arXiv240915714I}
{Ignace}, R.; {Scowen}, P.
\newblock {The Polstar UV Spectropolarimetry Mission}.
\newblock {\em arXiv e-prints} {\bf 2024}, p. arXiv:2409.15714,
  \href{http://arxiv.org/abs/2409.15714}{{\normalfont
  [arXiv:astro-ph.IM/2409.15714]}}.
\newblock {\url{https://doi.org/10.48550/arXiv.2409.15714}}.

\bibitem[{Lewis} et~al.(2022){Lewis}, {Bailey}, {Cotton}, {Howarth},
  {Kedziora-Chudczer}, and {van Leeuwen}]{2022MNRAS.513.1129L}
{Lewis}, F.; {Bailey}, J.; {Cotton}, D.V.; {Howarth}, I.D.;
  {Kedziora-Chudczer}, L.; {van Leeuwen}, F.
\newblock {A study of the F-giant star {\ensuremath{\theta}} Scorpii A: a
  post-merger rapid rotator?}
\newblock {\em \mnras} {\bf 2022}, {\em 513},~1129--1140,
  \href{http://arxiv.org/abs/2204.02719}{{\normalfont
  [arXiv:astro-ph.SR/2204.02719]}}.
\newblock {\url{https://doi.org/10.1093/mnras/stac991}}.

\bibitem[{Meade} et~al.(2012){Meade}, {Whitney}, {Babler}, {Nordsieck},
  {Bjorkman}, and {Wisniewski}]{2012AIPC.1429..226M}
{Meade}, M.R.; {Whitney}, B.A.; {Babler}, B.L.; {Nordsieck}, K.H.; {Bjorkman},
  K.S.; {Wisniewski}, J.P.
\newblock {HPOL: World's largest database of optical spectropolarimetry}.
\newblock In Proceedings of the Stellar Polarimetry: from Birth to Death;
  {Hoffman}, J.L.; {Bjorkman}, J.; {Whitney}, B., Eds. AIP,  2012, Vol. 1429,
  {\em American Institute of Physics Conference Series}, pp. 226--229.
\newblock {\url{https://doi.org/10.1063/1.3701930}}.

\bibitem[{Schulte-Ladbeck} et~al.(1990){Schulte-Ladbeck}, {Nordsieck}, {Nook},
  {Magalhaes}, {Taylor}, {Bjorkman}, and {Anderson}]{1990ApJ...365L..19S}
{Schulte-Ladbeck}, R.E.; {Nordsieck}, K.H.; {Nook}, M.A.; {Magalhaes}, A.M.;
  {Taylor}, M.; {Bjorkman}, K.S.; {Anderson}, C.M.
\newblock {A Rotating, Expanding Disk in the Wolf-Rayet Star EZ Canis Majoris?}
\newblock {\em \apjl} {\bf 1990}, {\em 365},~L19.
\newblock {\url{https://doi.org/10.1086/185878}}.

\bibitem[{Schulte-Ladbeck} et~al.(1992){Schulte-Ladbeck}, {Nordsieck},
  {Taylor}, {Bjorkman}, {Magalhaes}, and {Wolff}]{1992ApJ...387..347S}
{Schulte-Ladbeck}, R.F.; {Nordsieck}, K.H.; {Taylor}, M.; {Bjorkman}, K.S.;
  {Magalhaes}, A.M.; {Wolff}, M.J.
\newblock {The Wind Geometry of the Wolf-Rayet Star HD 191765}.
\newblock {\em \apj} {\bf 1992}, {\em 387},~347.
\newblock {\url{https://doi.org/10.1086/171087}}.

\bibitem[{{\"O}hman}(1946)]{1946ApJ...104..460O}
{{\"O}hman}, Y.
\newblock {On the Possibility of Tracing Polarization Effects in the Rotational
  Profiles of Early-Type Stars.}
\newblock {\em ApJ} {\bf 1946}, {\em 104},~460.
\newblock {\url{https://doi.org/10.1086/144879}}.

\bibitem[{Bailey} et~al.(2024){Bailey}, {Lewis}, {Howarth}, {Cotton},
  {Marshall}, and {Kedziora-Chudczer}]{2024arXiv240711352B}
{Bailey}, J.; {Lewis}, F.; {Howarth}, I.D.; {Cotton}, D.V.; {Marshall}, J.P.;
  {Kedziora-Chudczer}, L.
\newblock {Epsilon Sagittarii: An Extreme Rapid Rotator with a Decretion Disk}.
\newblock {\em arXiv e-prints} {\bf 2024}, p. arXiv:2407.11352,
  \href{http://arxiv.org/abs/2407.11352}{{\normalfont
  [arXiv:astro-ph.SR/2407.11352]}}.
\newblock {\url{https://doi.org/10.48550/arXiv.2407.11352}}.

\bibitem[{Collins} and {Cranmer}(1991)]{1991MNRAS.253..167C}
{Collins}, II, G.W.; {Cranmer}, S.R.
\newblock {Rotationally induced polarization in pure absorption spectral
  lines}.
\newblock {\em \mnras} {\bf 1991}, {\em 253},~167--174.
\newblock {\url{https://doi.org/10.1093/mnras/253.1.167}}.

\bibitem[{Trujillo Bueno}(2001)]{2001ASPC..236..161T}
{Trujillo Bueno}, J.
\newblock {Atomic Polarization and the Hanle Effect}.
\newblock In Proceedings of the Advanced Solar Polarimetry -- Theory,
  Observation, and Instrumentation; {Sigwarth}, M., Ed.,  2001, Vol. 236, {\em
  Astronomical Society of the Pacific Conference Series}, p. 161,
  \href{http://arxiv.org/abs/astro-ph/0202328}{{\normalfont
  [arXiv:astro-ph/astro-ph/0202328]}}.
\newblock {\url{https://doi.org/10.48550/arXiv.astro-ph/0202328}}.

\bibitem[{Stenflo}(2013)]{2013A&ARv..21...66S}
{Stenflo}, J.O.
\newblock {Solar magnetic fields as revealed by Stokes polarimetry}.
\newblock {\em A\&ARv} {\bf 2013}, {\em 21},~66,
  \href{http://arxiv.org/abs/1309.5454}{{\normalfont
  [arXiv:astro-ph.SR/1309.5454]}}.
\newblock {\url{https://doi.org/10.1007/s00159-013-0066-3}}.

\bibitem[{Ignace} et~al.(1997){Ignace}, {Nordsieck}, and
  {Cassinelli}]{1997ApJ...486..550I}
{Ignace}, R.; {Nordsieck}, K.H.; {Cassinelli}, J.P.
\newblock {The Hanle Effect as a Diagnostic of Magnetic Fields in Stellar
  Envelopes. I. Theoretical Results for Integrated Line Profiles}.
\newblock {\em ApJ} {\bf 1997}, {\em 486},~550--570.
\newblock {\url{https://doi.org/10.1086/304512}}.

\bibitem[{Manso Sainz} and {Mart{\'\i}nez
  Gonz{\'a}lez}(2012)]{2012ApJ...760....7M}
{Manso Sainz}, R.; {Mart{\'\i}nez Gonz{\'a}lez}, M.J.
\newblock {Hanle Effect for Stellar Dipoles and Quadrupoles}.
\newblock {\em \apj} {\bf 2012}, {\em 760},~7,
  \href{http://arxiv.org/abs/1209.6187}{{\normalfont
  [arXiv:astro-ph.SR/1209.6187]}}.
\newblock {\url{https://doi.org/10.1088/0004-637X/760/1/7}}.

\bibitem[{Folsom} et~al.(2022){Folsom}, {Ignace}, {Erba}, {Casini}, {del Pino
  Alem{\'a}n}, {Gayley}, {Hobbs}, {Manso Sainz}, {Neiner}, {Petit}, {Shultz},
  and {Wade}]{2022Ap&SS.367..125F}
{Folsom}, C.P.; {Ignace}, R.; {Erba}, C.; {Casini}, R.; {del Pino Alem{\'a}n},
  T.; {Gayley}, K.; {Hobbs}, K.; {Manso Sainz}, R.; {Neiner}, C.; {Petit}, V.;
  et~al.
\newblock {Ultraviolet spectropolarimetry: investigating stellar magnetic field
  diagnostics}.
\newblock {\em Ap\&SS} {\bf 2022}, {\em 367},~125,
  \href{http://arxiv.org/abs/2207.01865}{{\normalfont
  [arXiv:astro-ph.SR/2207.01865]}}.
\newblock {\url{https://doi.org/10.1007/s10509-022-04140-8}}.

\bibitem[{Neiner} et~al.(2015){Neiner}, {Mathis}, {Alecian}, {Emeriau},
  {Grunhut}, {BinaMIcS}, and {MiMeS Collaborations}]{2015IAUS..305...61N}
{Neiner}, C.; {Mathis}, S.; {Alecian}, E.; {Emeriau}, C.; {Grunhut}, J.;
  {BinaMIcS}.; {MiMeS Collaborations}.
\newblock {The origin of magnetic fields in hot stars}.
\newblock In Proceedings of the Polarimetry; {Nagendra}, K.N.; {Bagnulo}, S.;
  {Centeno}, R.; {Jes{\'u}s Mart{\'\i}nez Gonz{\'a}lez}, M., Eds.,  2015, Vol.
  305, {\em IAU Symposium}, pp. 61--66,
  \href{http://arxiv.org/abs/1502.00226}{{\normalfont
  [arXiv:astro-ph.SR/1502.00226]}}.
\newblock {\url{https://doi.org/10.1017/S1743921315004524}}.

\bibitem[{ud-Doula} and {Owocki}(2002)]{2002ApJ...576..413U}
{ud-Doula}, A.; {Owocki}, S.P.
\newblock {Dynamical Simulations of Magnetically Channeled Line-driven Stellar
  Winds. I. Isothermal, Nonrotating, Radially Driven Flow}.
\newblock {\em \apj} {\bf 2002}, {\em 576},~413--428,
  \href{http://arxiv.org/abs/astro-ph/0201195}{{\normalfont
  [arXiv:astro-ph/astro-ph/0201195]}}.
\newblock {\url{https://doi.org/10.1086/341543}}.

\bibitem[{Petit} et~al.(2013){Petit}, {Owocki}, {Wade}, {Cohen}, {Sundqvist},
  {Gagn{\'e}}, {Ma{\'\i}z Apell{\'a}niz}, {Oksala}, {Bohlender}, {Rivinius},
  {Henrichs}, {Alecian}, {Townsend}, {ud-Doula}, and {MiMeS
  Collaboration}]{2013MNRAS.429..398P}
{Petit}, V.; {Owocki}, S.P.; {Wade}, G.A.; {Cohen}, D.H.; {Sundqvist}, J.O.;
  {Gagn{\'e}}, M.; {Ma{\'\i}z Apell{\'a}niz}, J.; {Oksala}, M.E.; {Bohlender},
  D.A.; {Rivinius}, T.;  et~al.
\newblock {A magnetic confinement versus rotation classification of
  massive-star magnetospheres}.
\newblock {\em \mnras} {\bf 2013}, {\em 429},~398--422,
  \href{http://arxiv.org/abs/1211.0282}{{\normalfont
  [arXiv:astro-ph.SR/1211.0282]}}.
\newblock {\url{https://doi.org/10.1093/mnras/sts344}}.

\bibitem[{Owocki} et~al.(2020){Owocki}, {Shultz}, {ud-Doula}, {Sundqvist},
  {Townsend}, and {Cranmer}]{2020MNRAS.499.5366O}
{Owocki}, S.P.; {Shultz}, M.E.; {ud-Doula}, A.; {Sundqvist}, J.O.; {Townsend},
  R.H.D.; {Cranmer}, S.R.
\newblock {How the breakout-limited mass in B-star centrifugal magnetospheres
  controls their circumstellar H {\ensuremath{\alpha}} emission}.
\newblock {\em \mnras} {\bf 2020}, {\em 499},~5366--5378,
  \href{http://arxiv.org/abs/2009.12359}{{\normalfont
  [arXiv:astro-ph.SR/2009.12359]}}.
\newblock {\url{https://doi.org/10.1093/mnras/staa2325}}.

\bibitem[{Owocki} et~al.(2022){Owocki}, {Shultz}, {ud-Doula}, {Chandra}, {Das},
  and {Leto}]{2022MNRAS.513.1449O}
{Owocki}, S.P.; {Shultz}, M.E.; {ud-Doula}, A.; {Chandra}, P.; {Das}, B.;
  {Leto}, P.
\newblock {Centrifugal breakout reconnection as the electron acceleration
  mechanism powering the radio magnetospheres of early-type stars}.
\newblock {\em \mnras} {\bf 2022}, {\em 513},~1449--1458,
  \href{http://arxiv.org/abs/2202.05449}{{\normalfont
  [arXiv:astro-ph.SR/2202.05449]}}.
\newblock {\url{https://doi.org/10.1093/mnras/stac341}}.

\bibitem[{Erba} et~al.(2021){Erba}, {David-Uraz}, {Petit}, {Hennicker},
  {Fletcher}, {Fullerton}, {Naz{\'e}}, {Sundqvist}, and
  {ud-Doula}]{2021MNRAS.506.5373E}
{Erba}, C.; {David-Uraz}, A.; {Petit}, V.; {Hennicker}, L.; {Fletcher}, C.;
  {Fullerton}, A.W.; {Naz{\'e}}, Y.; {Sundqvist}, J.; {ud-Doula}, A.
\newblock {Ultraviolet line profiles of slowly rotating massive star winds
  using the 'analytic dynamical magnetosphere' formalism}.
\newblock {\em \mnras} {\bf 2021}, {\em 506},~5373--5388,
  \href{http://arxiv.org/abs/2106.13676}{{\normalfont
  [arXiv:astro-ph.SR/2106.13676]}}.
\newblock {\url{https://doi.org/10.1093/mnras/stab1853}}.

\bibitem[{Townsend} and {Owocki}(2005)]{2005MNRAS.357..251T}
{Townsend}, R.H.D.; {Owocki}, S.P.
\newblock {A rigidly rotating magnetosphere model for circumstellar emission
  from magnetic OB stars}.
\newblock {\em \mnras} {\bf 2005}, {\em 357},~251--264,
  \href{http://arxiv.org/abs/astro-ph/0408565}{{\normalfont
  [arXiv:astro-ph/astro-ph/0408565]}}.
\newblock {\url{https://doi.org/10.1111/j.1365-2966.2005.08642.x}}.

\bibitem[{Sundqvist} et~al.(2012){Sundqvist}, {ud-Doula}, {Owocki}, {Townsend},
  {Howarth}, and {Wade}]{2012MNRAS.423L..21S}
{Sundqvist}, J.O.; {ud-Doula}, A.; {Owocki}, S.P.; {Townsend}, R.H.D.;
  {Howarth}, I.D.; {Wade}, G.A.
\newblock {A dynamical magnetosphere model for periodic H{\ensuremath{\alpha}}
  emission from the slowly rotating magnetic O star HD 191612}.
\newblock {\em \mnras} {\bf 2012}, {\em 423},~L21--L25,
  \href{http://arxiv.org/abs/1203.1050}{{\normalfont
  [arXiv:astro-ph.SR/1203.1050]}}.
\newblock {\url{https://doi.org/10.1111/j.1745-3933.2012.01248.x}}.

\bibitem[{Owocki} et~al.(2016){Owocki}, {ud-Doula}, {Sundqvist}, {Petit},
  {Cohen}, and {Townsend}]{2016MNRAS.462.3830O}
{Owocki}, S.P.; {ud-Doula}, A.; {Sundqvist}, J.O.; {Petit}, V.; {Cohen}, D.H.;
  {Townsend}, R.H.D.
\newblock {An `analytic dynamical magnetosphere' formalism for X-ray and
  optical emission from slowly rotating magnetic massive stars}.
\newblock {\em \mnras} {\bf 2016}, {\em 462},~3830--3844,
  \href{http://arxiv.org/abs/1607.08568}{{\normalfont
  [arXiv:astro-ph.SR/1607.08568]}}.
\newblock {\url{https://doi.org/10.1093/mnras/stw1894}}.

\bibitem[{Andersson} et~al.(2015){Andersson}, {Lazarian}, and
  {Vaillancourt}]{2015ARA&A..53..501A}
{Andersson}, B.G.; {Lazarian}, A.; {Vaillancourt}, J.E.
\newblock {Interstellar Dust Grain Alignment}.
\newblock {\em \araa} {\bf 2015}, {\em 53},~501--539.
\newblock {\url{https://doi.org/10.1146/annurev-astro-082214-122414}}.

\bibitem[{Serkowski} et~al.(1975){Serkowski}, {Mathewson}, and
  {Ford}]{1975ApJ...196..261S}
{Serkowski}, K.; {Mathewson}, D.S.; {Ford}, V.L.
\newblock {Wavelength dependence of interstellar polarization and ratio of
  total to selective extinction.}
\newblock {\em ApJ} {\bf 1975}, {\em 196},~261--290.
\newblock {\url{https://doi.org/10.1086/153410}}.

\bibitem[{Whittet} et~al.(1992){Whittet}, {Martin}, {Hough}, {Rouse}, {Bailey},
  and {Axon}]{1992ApJ...386..562W}
{Whittet}, D.C.B.; {Martin}, P.G.; {Hough}, J.H.; {Rouse}, M.F.; {Bailey},
  J.A.; {Axon}, D.J.
\newblock {Systematic Variations in the Wavelength Dependence of Interstellar
  Linear Polarization}.
\newblock {\em ApJ} {\bf 1992}, {\em 386},~562.
\newblock {\url{https://doi.org/10.1086/171039}}.

\bibitem[{Ignace} and {Scowen}(2024)]{2024BSRSL..93..156I}
{Ignace}, R.; {Scowen}, P.
\newblock {The Polstar UV spectropolarimetry mission}.
\newblock {\em Bulletin de la Societe Royale des Sciences de Liege} {\bf 2024},
  {\em 93},~156--172,  \href{http://arxiv.org/abs/2409.15714}{{\normalfont
  [arXiv:astro-ph.IM/2409.15714]}}.
\newblock {\url{https://doi.org/10.25518/0037-9565.12308}}.

\bibitem[{St-Louis} and {Polstar consortium}(2024)]{2024IAUS..361..633S}
{St-Louis}, N.; {Polstar consortium}.
\newblock {Polstar : a FUV Spectropolarimetry Mission}.
\newblock In Proceedings of the IAU Symposium; {Mackey}, J.; {Vink}, J.S.;
  {St-Louis}, N., Eds.,  2024, Vol. 361, {\em IAU Symposium}, pp. 633--635.
\newblock {\url{https://doi.org/10.1017/S1743921322003167}}.

\bibitem[{Girardot} et~al.(2024){Girardot}, {Neiner}, and
  {Reess}]{2024sf2a.conf..457G}
{Girardot}, A.; {Neiner}, C.; {Reess}, J.M.
\newblock {UV spectropolarimetry with CASSTOR, Polstar, and Pollux}.
\newblock In Proceedings of the SF2A-2024: Proceedings of the Annual meeting of
  the French Society of Astronomy and Astrophysics. Eds.: M. B{\'e}thermin;
  {B{\'e}thermin}, M.; {Bailli{\'e}}, K.; {Lagarde}, N.; {Malzac}, J.;
  {Ouazzani}, R.M.; {Richard}, J.; {Venot}, O.; {Siebert}, A., Eds.,  2024, pp.
  457--460.

\bibitem[{Nordsieck} et~al.(1994){Nordsieck}, {Code}, {Anderson}, {Meade},
  {Babler}, {Michalski}, {Pfeifer}, and {Jones}]{1994SPIE.2010....2N}
{Nordsieck}, K.H.; {Code}, A.D.; {Anderson}, C.M.; {Meade}, M.R.; {Babler}, B.;
  {Michalski}, D.E.; {Pfeifer}, R.H.; {Jones}, T.E.
\newblock {Exploring ultraviolet astronomical polarimetry: results from the
  Wisconsin Ultraviolet Photo-Polarimeter Experiment (WUPPE)}.
\newblock In Proceedings of the X-Ray and Ultraviolet Polarimetry; {Fineschi},
  S., Ed.,  1994, Vol. 2010, {\em Society of Photo-Optical Instrumentation
  Engineers (SPIE) Conference Series}, pp. 2--11.
\newblock {\url{https://doi.org/10.1117/12.168568}}.

\bibitem[{Boggess} et~al.(1978){Boggess}, {Carr}, {Evans}, {Fischel},
  {Freeman}, {Fuechsel}, {Klinglesmith}, {Krueger}, {Longanecker}, and
  {Moore}]{1978Natur.275..372B}
{Boggess}, A.; {Carr}, F.A.; {Evans}, D.C.; {Fischel}, D.; {Freeman}, H.R.;
  {Fuechsel}, C.F.; {Klinglesmith}, D.A.; {Krueger}, V.L.; {Longanecker}, G.W.;
  {Moore}, J.V.
\newblock {The IUE spacecraft and instrumentation}.
\newblock {\em \nat} {\bf 1978}, {\em 275},~372--377.
\newblock {\url{https://doi.org/10.1038/275372a0}}.

\bibitem[{Massa} et~al.(1995){Massa}, {Fullerton}, {Nichols}, {Owocki},
  {Prinja}, {St-Louis}, {Willis}, {Altner}, {Bolton}, {Cassinelli}, {Cohen},
  {Cooper}, {Feldmeier}, {Gayley}, {Harries}, {Heap}, {Henriksen}, {Howarth},
  {Hubeny}, {Kambe}, {Kaper}, {Koenigsberger}, {Marchenko}, {McCandliss},
  {Moffat}, {Nugis}, {Puls}, {Robert}, {Schulte-Ladbeck}, {Smith}, {Smith},
  {Waldron}, and {White}]{1995ApJ...452L..53M}
{Massa}, D.; {Fullerton}, A.W.; {Nichols}, J.S.; {Owocki}, S.P.; {Prinja},
  R.K.; {St-Louis}, N.; {Willis}, A.J.; {Altner}, B.; {Bolton}, C.T.;
  {Cassinelli}, J.P.;  et~al.
\newblock {The IUE MEGA Campaign: Wind Variability and Rotation in Early-Type
  Stars}.
\newblock {\em \apjl} {\bf 1995}, {\em 452},~L53.
\newblock {\url{https://doi.org/10.1086/309707}}.

\end{thebibliography}

\PublishersNote{}
\end{adjustwidth}
\end{document}